\documentclass[twocolumn,trackchanges]{aastex62}
\graphicspath{{./}}
\usepackage{color}
\usepackage{amsmath}
\usepackage{array}
\usepackage{xfrac}
\usepackage[bottom]{footmisc}
\usepackage{lipsum}
\usepackage{tablefootnote}
\usepackage{textcomp}
\usepackage{booktabs}
\usepackage{listings}
\usepackage{graphicx}
\usepackage{float}

\definecolor{codegreen}{rgb}{0,0.6,0}
\definecolor{codegray}{rgb}{0.5,0.5,0.5}
\definecolor{codepurple}{rgb}{0.58,0,0.82}
\definecolor{backcolour}{rgb}{0.95,0.95,0.92}

\shorttitle{A-LIST}
\shortauthors{Ashok et al.}

\begin{document}

\title{The APOGEE Library of Infrared SSP Templates (A-LIST): \\ High-Resolution Simple Stellar Population Spectral Models in the $H$-Band}

\author{Aishwarya Ashok}
\affiliation{Department of Physics and Astronomy, University of Utah, Salt Lake City, UT 84112, USA}

\author{Gail Zasowski}
\affiliation{Department of Physics and Astronomy, University of Utah, Salt Lake City, UT 84112, USA}

\author{Anil Seth}
\affiliation{Department of Physics and Astronomy, University of Utah, Salt Lake City, UT 84112, USA}

\author{Sten Hasselquist}
\altaffiliation{NSF Postdoctoral Fellow}
\affiliation{Department of Physics and Astronomy, University of Utah, Salt Lake City, UT 84112, USA}

\author{Galen Bergsten}
\affiliation{Department of Physics and Astronomy, University of Utah, Salt Lake City, UT 84112, USA}

\author{Olivia Cooper}
\affiliation{Department of Astronomy, The University of Texas at Austin, 2515 Speedway Boulevard Stop C1400, Austin, TX 78712, USA}

\author{Nicholas Boardman}
\affiliation{Department of Physics and Astronomy, University of Utah, Salt Lake City, UT 84112, USA}

\author {Dmitry Bizyaev}
\affiliation{Apache Point Observatory and New Mexico State University, P.O. Box 59, Sunspot, NM, 88349-0059, USA}
\affiliation{Sternberg Astronomical Institute, Moscow State University, Moscow, Russia}

\author{Sofia Meneses Goytia}
\affiliation{Department of Physics, Astrophysics Research Group, University of Surrey, Guildford, Surrey, GU2 7X, UK}

\author{D. A. García-Hernández}
\affiliation{Instituto de Astrofísica de Canarias (IAC), E-38205 La Laguna, Tenerife, Spain}
\affiliation{Universidad de La Laguna (ULL), Departamento de Astrofísica, E-38206, La Laguna, Tenerife, Spain}

\author{Alexandre Roman-Lopes}
\affiliation{Departamento de Física, Facultad de Ciencias, Universidad de La Serena, Cisternas 1200, La Serena, Chile}

\begin{abstract} \label{abstract}
Integrated light spectroscopy from galaxies can be used to study the stellar populations that cannot be resolved into individual stars. This analysis relies on stellar population synthesis (SPS) techniques to study the formation history and structure of galaxies.  However, the spectral templates available for SPS are limited, especially in the near-infrared. We present A-LIST (APOGEE Library of Infrared SSP Templates), a new set of high-resolution, near-IR SSP spectral templates spanning a wide range of ages (2--12~Gyr), metallicities ($\rm -2.2 < [M/H] < +0.4$) and $\alpha$ abundances ($\rm -0.2 < [\alpha/M] < +0.4$). This set of SSP templates is the highest resolution ($R\sim22500$) available in the near infrared, and the first such based on an empirical stellar library. Our models are generated using spectra of $\sim$300,000 stars spread across the Milky Way, with a wide range of metallicities and abundances, from the APOGEE survey. We show that our model spectra provide accurate fits to M31 globular cluster spectra taken with APOGEE, with best-fit metallicities agreeing with those of previous estimates to within $\sim$0.1~dex.  We also compare these model spectra to lower-resolution E-MILES models and demonstrate that we recover the ages of these models to within $\sim$1.5 Gyr. This library is available in \url{https://github.com/aishashok/ALIST-library}.
\end{abstract}


\section{Introduction} 
\label{intro}
Observing, modeling and analysis of individual stars in the visible spectrum has formed the foundation for our understanding of stellar evolution. In turn, our understanding of galaxies today: their star formation rates, evolutionary histories, and stellar masses, is almost entirely based on our knowledge of stellar evolution.
Resolved stellar populations allow us to measure the colors, luminosities and compositions of individual stars, provides useful information about the history and structure of their galaxies. Resolved population studies are possible only in the Milky Way \citep[e.g.,][]{Ruiz-lara2020} as well as nearby spiral and dwarf galaxies \citep[e.g,][]{Williams2009, Weisz2011, Lewis2015}.
In more distant galaxies, integrated spectroscopy has been an important method to study their unresolved stellar content. The integrated spectrum of a population reflects its star formation histories and chemical enrichment. Using spectral lines to study the physical properties of stars and stellar populations has been known as early as 1960s and has improved ever since
\citep{Spinrad1971, Mould1978, Chavez1996, Parikh2019}. 

Earlier work studying galaxy stellar populations focused primarily on line index measurements \citep[e.g.][]{Worthey1994, Trager2000, Thomas2003, Gallazzi2005, Schiavon2007,Caldwell2011} in the optical part of the spectrum as well as in the near-infrared \citep[NIR; e.g.][]{Vazdekis1996, Boker1999, Cenarro2001, Forster2003, Cenarro2009, MacArthur2010, Vazdekis2010, Onken2014, Riffel2019}. In recent years, full spectral fitting  of integrated light spectra to model spectra has opened a new window to study in detail the physical, chemical and evolutionary phases of galaxy stellar populations \citep[e.g.][]{Perez2013, Pace2019, Boardman2020}. Some of the most widely used spectral fitting software for this technique are FIREFLY \citep{wilkinson2017}, STECKMAP \citep{ocvirk2006}, VESPA \citep{tojeiro2007}, pPXF \citep{Cappellari2009}, ULySS\citep{Koleva2009}, STARLIGHT \citep{cidfernandes2011}, and Pipe3D \citep{sanchez2016a, sanchez2016b}. Most of the pixel fitting stellar population studies to date have used optical spectroscopy to derive information about galaxy formation histories \citep[e.g.][]{heavens2004, McDermid2015}. .The NIR part of the spectrum offers some advantages  over the optical due to the reduced affects of dust extinction and the sensitivity to cool stars. Over the past decade or so, a large number of studies have studied stellar populations using the NIR \citep[e.g.][]{ silva2008, cesetti2009, marmol2008, Sakari2016, labarbera2016, labarbera2017, rock2017,  Baldwin2018,  Dahmer-Hahn2018, Martins2019, Coelho2020}.

These direct spectral fitting techniques use the Stellar Population Synthesis (SPS) technique, where spectra are modeled with a combination of synthetic populations to disentangle their age and chemical abundances. The foundation of this method is the Simple Stellar Population (SSP) model where all the stars are coeval and share the same chemical composition. Creating model stellar populations has progressed greatly in the last three decades from the trial and error technique \citep[e.g.][]{Tinsley1968, Bruzal2003, Maraston2005, Vazdekis2010, Rock2016, Conroy2018}.

Given the wide range of information we want to gain from SPS fitting: the star formation history and rate, initial mass function, chemical enrichment, metallicity etc., creating optimal and flexible SSP models remains a major challenge in modern evolutionary studies.
There are 2 ways these SSPs can be constructed
(i) Fuel Consumption Based approach  \citep[e.g.][]{Maraston2005} and (ii) The Isochrone Synthesis approach which we use in this paper. With the Isochrone Synthesis Method, we can construct the SSPs as having an Initial Mass Function along an isochrone and a stellar library \citep[e.g.][]{Bruzal2003}.

The theoretical stellar libraries \citep[e.g. MARCS library,][]{Gustafsson2008} are generated using a desired range of parameters that cannot be obtained from observations at any range of wavelength, while the empirical libraries \citep[e.g.,ELODIE library, X-shooter Library (R$\sim$10,000)][]{Prugniel2007, Gonneau2020} are based on real observational data. Large spectroscopic surveys like APOGEE \citep{Majewski2017}, GALAH \citep{DeSilva2015}, etc. can provide us with tens of thousands of Milky Way star properties that can be used to model stellar populations.
To study the stellar absorption features in old populations, empirical libraries are preferred \citep{Maraston2009, vazdekis2016}, while for young and bright populations, theoretical libraries are preferred \citep{Maraston2005, Leitherer2014}. \citet{Martins2019, Coelho2020} explain in detail how these two types of libraries are used.

The models we present here address two specific deficits in existing SPS models.  First, most models don't consider varying $\alpha$ abundances, and $\alpha$ abundances are typically still derived using spectral indices rather than pixel fitting to SPS models \citep[e.g.][]{Schiavon2012, Janz2016, JohnstonEvelyn2020}. Second, there are no SPS models based on empirical stellar libraries in the NIR available at high spectral resolution. The only high spectral resolution NIR models (R$\sim$20,000) are the version of the \citet{Maraston2011} models created using the MARCS theoretical stellar spectra library. 
See \citet{Baldwin2018} for how these models compare with other available NIR models and data.
Existing models based on empirical stellar libraries available in the NIR \citep{Bruzal2003, Maraston2011, Meneses-Goytia2015, vazdekis2016, Rock2016,Conroy2018} use the  \citet{Pickles1998} stellar library and the IRTF library \citep{Cushing2005,Rayner2009}. These are both low spectral resolution, with the IRTF library having the higher resolution of R$\sim$2000.

This spectral resolution is lower than that of many current and planned spectrographs: Gemini/NIFS \citep{NIFS}, VLT/SINFONI \citep{SINFONI}, GEMINI/GNIRS \citep{GNIRS}, ESO/KMOS \citep{KMOS}, VLT/MOONS \citep{MOONS}, Keck/MOSFIRE \citep{MOSFIRE}, JWST/NIRSPEC \citep{NIRSPEC}. For data taken from these spectrographs, using the current available models based on empirical stellar libraries requires degrading the data resolution and losing information while modeling. 

In this paper, we present A-LIST (APOGEE Library of Infrared SSP Templates), a new library of high resolution SSP spectral models generated using a new empirical stellar library based on Apache Point Observatory Galaxy Evolution Experiment \citep[APOGEE;][]{Majewski2017} data from the Sloan Digital Sky Survey-IV \citep[SDSS-IV;][]{Blanton2017}. The SSP models range in age, [M/H] and [$\alpha$/M]. This is a first of its kind spectral library in the H-band region having a high spectral resolution (R$\sim$ 22500). This library can be used to model, for example, (i) any near IR integrated light spectra having a spectral resolution greater than the current available SSP models; (ii) $\alpha$ enhancement and kinematics of globular clusters (GCs) and/or early-type galaxies; and (iii) nearby galaxy data (M31, M33) available from APOGEE \citep[][Beaton et al. in prep]{Zasowski2013}.
Because APOGEE spectra are not precisely flux calibrated,  A-LIST spectral models are primarily useful for integrated light spectroscopic studies.
\begin{figure*}
\includegraphics[trim=2cm 0cm 2cm 1cm, clip, width=\textwidth]{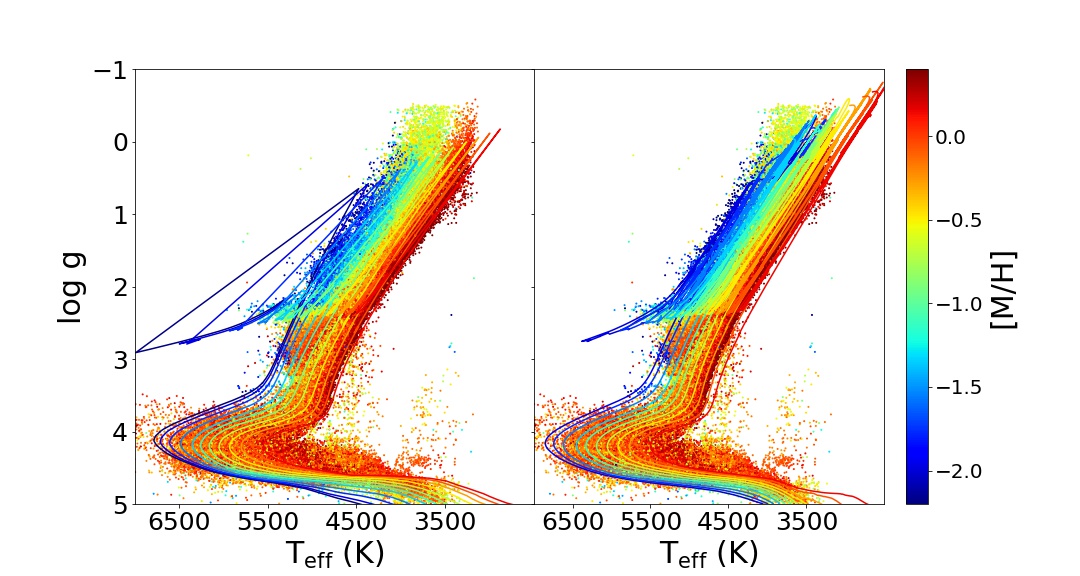}
\caption{Isochrones (lines) plotted over the empirical library of APOGEE stars (points).  Both are colored by their metallicity.  The isochrones plotted are  for a 10 Gyr and solar [$\alpha$/M] population. The left panel shows the Padova isochrones, and the right panel shows the same for the MIST isochrones.}
\label{fig:isochrones}
\end{figure*}

We discuss the required ingredients and our selections in Section \ref{ingredients}. In Section \ref{generation}, we discuss how we generate the Simple Stellar Population models in detail. In Section \ref{description}, we present our spectral models. Section \ref{validation}, we present our validation of these models by reproducing properties of M31 GCs and comparisons to previously studied models. The summary along with future work is described in Section \ref{summary}.

\section{Simple Stellar Population Model Ingredients} 
\label{ingredients}
In this section, we outline the main ingredients, including the isochrone sets (Section~\ref{isochrones}) and the stellar library (Section~\ref{apogeestars}).

\subsection{Stellar Evolution and Isochrones} 
\label{isochrones}
One of the main ingredients to generate an SSP model is an isochrone or a set of stellar evolutionary tracks. 

For our models, we use two different isochrone sets:
\begin{itemize}
\item PARSEC isochrones: the latest Padova isochrones and stellar evolutionary tracks \citep{Bressan2012,Chen2014,Chen2015,Tang2014,Marigo2017,Pastorelli2019} are computed spanning a wide range in ages and metallicities. The library of isochrones includes stars with initial masses $M=0.15-100$~M$_\odot$, $\rm \log(age)=6.6-10.2$, and  metallicities $\rm [M/H] = -2.2-+0.5$. The PARSEC stellar tracks range in initial metal content $(z_{i})$ from 0.0001 to 0.06, with 15 values in the grid, and range in mass from 0.1 to 350 M$_\odot$ with $\sim$120 different mass values for each metallicity \citep{Marigo2017}.

\item MIST evolutionary tracks:
MESA Isochrones and Stellar Tracks \citep[MIST;][]{Dotter2016,Choi2016,Paxton2011,Paxton2013,Paxton2015} is another set of evolutionary track models that has an extensive range of mass ($M=0.1-300$~M$_\odot$), age ($\rm \log(age)=5-10.3$), and metallicity ($\rm -4 \leq [M/H] \leq +0.5$). The evolutionary tracks used to generate the isochrones range in mass from 0.1 to 300 M$_\odot$ with $>$ 100 models at different masses, and in [Fe/H] from -2.0 to +0.5 with 0.25 dex spacing  \citep{Choi2016}.
We use the models with $v/v_{\rm crit} = 0.0$, i.e., those models which do not include any stellar rotation parameters in them.
\end{itemize}
Both sets of models use interpolation of the evolutionary tracks to create isochrones at a given age and metallicity. The list of ages and metallicities used in our models is given in Section \ref{description}.

One of the main differences between the PARSEC and the MIST isochrones is the method used to interpolate the TP-AGB evolutionary tracks.  This results in the C-rich AGB stars in the MIST isochrones starting at a later point in their evolutionary track compared to those of PARSEC \citep{Marigo2017,Choi2016}. The C-rich AGB stars in MIST isochrones are hotter than those in PARSEC mainly because the MIST isochrones use molecular opacities derived for O-rich mixtures \citep{Marigo2017}. For a detailed description of how these two isochrone sets differ, see  \citet{Marigo2017, Choi2016,Cignoni2019}.

Figure \ref{fig:isochrones} shows the two isochrone sets plotted over the empirical library of APOGEE stars (Section~\ref{apogeestars}) colored by metallicity, with Padova isochrones on the left and MIST on the right.

The isochrones available in Padova and MIST have solar $\alpha$--abundances. 
To obtain the most accurate isochrones for our models with non-solar $\alpha$--abundance ($\rm [\alpha/M]$), we use the $\alpha$-enhanced metallicity equation \citep[from][chapter 8]{Salaris2005} for the solar-scaled isochrones with the metallicity as:  
\begin{equation}
\rm [M/H]_{\rm iso} = [M/H] + log(0.694 \times 10^{[\alpha/M]} + 0.306).
\end{equation}
where $\rm [M/H]$ is the solar scaled metallicity 
and $\rm [M/H]_{\rm iso}$ gives the effective metallicity of the isochrones used to account for the $\alpha$--enhancement.
We note that we use this $\rm [M/H]_{\rm iso}$ value only for selecting the isochrones of appropriate metallicity for a model with desired $\rm [\alpha/M]$ and solar-scaled $\rm[M/H]$.

\subsection{Empirical Stellar Library} \label{apogeestars}

\begin{figure*}
\advance\leftskip-1cm
\includegraphics[scale=0.4]{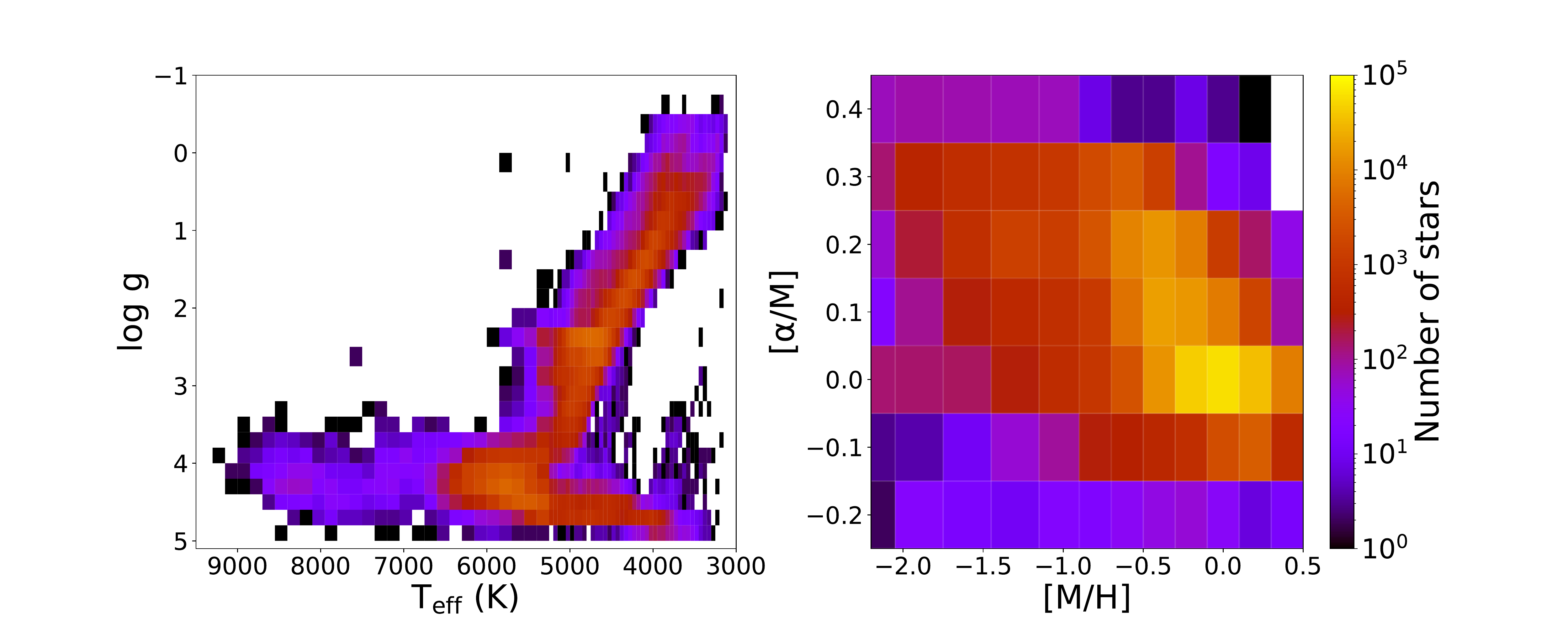}
\caption{The empirical APOGEE stellar library used to generate our SSP spectral models. The left panel shows a HR Diagram with the number of APOGEE stars selected per T$_\mathrm{eff}$ and surface gravity bin.  The right panel shows the number of selected APOGEE stars in [M/H] and [$\alpha$/M] space. These bins are defined in Section~\ref{paramgrid}.}
\label{apogeeparam}
\end{figure*}

The second main ingredient needed to generate an SSP model is the empirical stellar spectral library. One strength of A-LIST is the availability of several hundred thousand stellar spectra from the APOGEE survey that can be used as an empirical library.  
APOGEE \citep{Majewski2017} is a spectroscopic survey of the Milky Way and a component of the Sloan Digital Sky Survey \citep[SDSS-III and -IV;][]{Eisenstein2011,Blanton2017}

APOGEE samples stars that span across the Milky Way's bulge, disk and halo with a wide range of stellar parameters and abundances  \citep{Zasowski2017,Zasowski2013}. The data are collected with two 300-fiber spectrographs  \citep{Wilson2019} at the 2.5-meter Sloan Foundation Telescope at the Apache Point Observatory in New Mexico \citep{Gunn2006} and at the 2.5-meter duPont Telescope at Las Campanas Observatory in Chile \citep{Bowen1973}. The wavelength range of APOGEE spectra is 1.51--1.70$\mu$m, which is divided across three chips: ``blue'' chip at $1.51-1.581\mu$m, ``green'' chip at $1.585-1.644\mu$m, and ``red'' chip at $1.647-1.700\mu$m, with gaps between the detectors at $1.581-1.585\mu$m and $1.644-1.647\mu$m). APOGEE's average resolving power ($R$) is $\sim$ 22,500 based on a direct-measured FWHM of $\sim$0.7\AA, with 10-20\% variations seen across the wavelength and fiber \citep[see Section 6.2][]{Nidever2015}.

We use an internal APOGEE data release that is an increment from the latest public data release \citep[DR16;][]{Ahumada2019}, which has new stars added from the observing period through November 2019.
The incremental release applies the same DR16 pipeline \citep[][]{jonsson2020} to the additional data, which include both new targets and repeat visits for stars in DR16.

APOGEE data are first processed by the APOGEE data reduction pipeline \citep{Nidever2015}. The stellar parameters and elemental abundances are computed by the APOGEE Stellar Parameters and Chemical Abundances Pipeline \citep[ASPCAP;][]{Garcia2016}. For our empirical library, we use the calibrated ASPCAP parameters to select cool dwarfs and giants, and the uncalibrated parameters (``FPARAM'') to select hot dwarfs ($T{\rm eff} \geq 7000$~K).
See \citet{jonsson2020} and \citet{Holtzman2018} for a description of APOGEE's calibration procedures and data products.

To obtain a set of APOGEE stars with reliable ASPCAP measurements and high SNR we make the following cuts:
\begin{enumerate}
\item SNR: We remove all stars with median $\rm S/N$ per pixel $< 100$.

When creating the median spectrum for a given stellar parameter bin (Section~\ref{paramgrid}), we only consider pixels with an individual $\rm SNR \geq 150$, computed with the stellar error spectra.
\item VSCATTER: We remove all stars that have a radial velocity scatter greater than 1.5~km~s$^{-1}$. This cut reduces the impact of  spectroscopic binary stars on our sample.
\item VERR: We remove all stars that have a radial velocity error greater than 3~km~s$^{-1}$. This limit ensures that the selected stars are properly corrected to the rest frame.
\item ASPCAPFLAG bitmask: We remove all stars that are marked as STAR$\_$BAD (bit 23), TEFF$\_$BAD (bit 16), LOGG$\_$BAD (bit 17), or COLORTE$\_$WARN (bit 9).
\item STARFLAG bitmask: We remove all stars that are marked BRIGHT$\_$NEIGHBOR (bit 2), VERY$\_$BRIGHT$\_$NEIGHBOR (bit 3), PERSIST$\_$HIGH (bit 9), PERSIST$\_$MED (bit 10), PERSIST$\_$LOW (bit 11), SUSPECT$\_$RV$\_$COMBINATION (bit 16), or SUSPECT$\_$BROAD$\_$LINES (bit 17).
 \end{enumerate}

\begin{figure*}
\includegraphics[width=\textwidth]{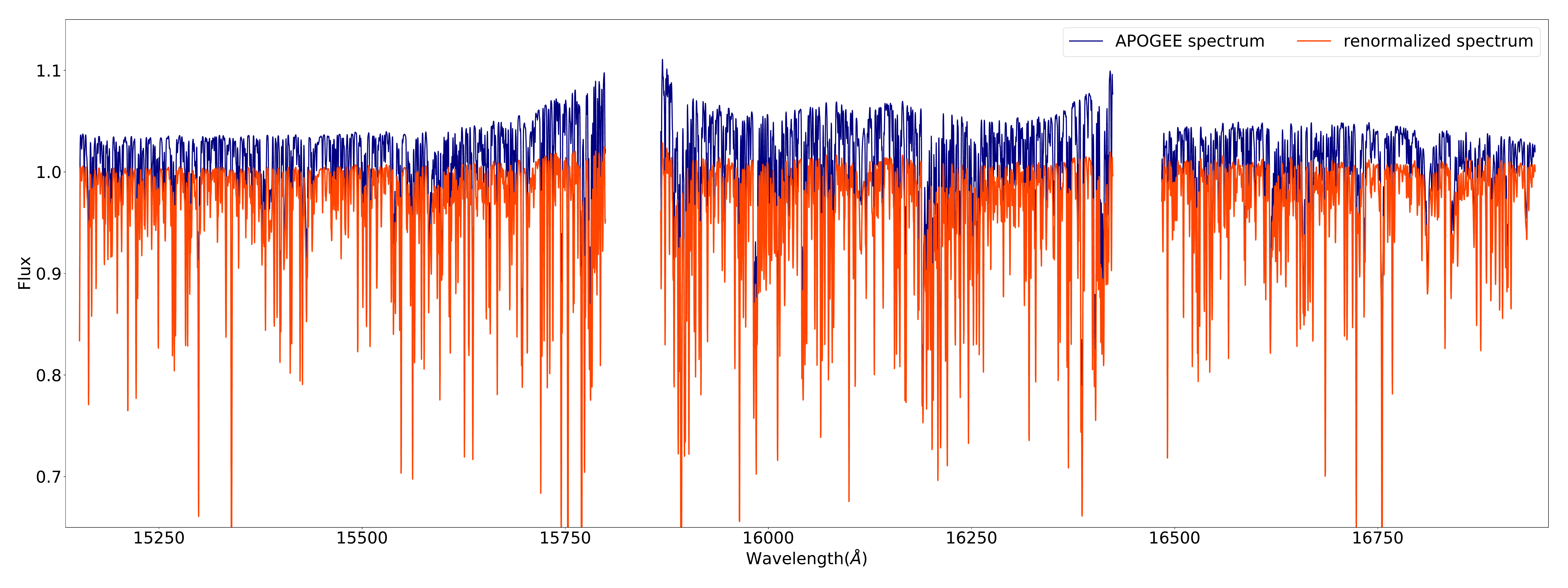}
\caption{Example of the re-normalization of a stellar spectrum. Blue denotes the pseudo-continuum normalized APOGEE spectrum, and orange denotes the renormalized spectrum we use in our library. The gaps at 15810\AA~and 16440\AA~correspond to gaps between APOGEE's detectors.}
\label{renorm}
\end{figure*}

Based on these cuts, we select 293,220 stars from the incremental DR16 data.
In this paper, we use [M/H] as our solar-scaled metallicity based on its definition in Section 6.5.1 of \citet{Majewski2017}.
The left panel of Figure~\ref{apogeeparam} shows an HR Diagram of the APOGEE stellar sample that we use as our empirical stellar library. The right panel of Figure~\ref{apogeeparam} shows the [M/H] and [$\alpha$/M] distribution of the APOGEE stars. Both the plots are colored by the number of stars in each bin. The axes on these two plots define the limits within which our models are generated.

We found it necessary to renormalize some of the pseudo-continuum normalized APOGEE spectra, due to systematic offsets in the pseudo-continuum determination, before combining the spectra. For all stars whose median normalized flux is greater than 1.04, and stars that show wing-like features (especially at the chip edges), we renormalize the spectra by fitting a 3rd, 4th, and 3rd order polynomial to the blue, green, and red chips of the spectra, respectively. An example of this is shown in Figure~\ref{renorm}. Here the blue line shows the APOGEE pseudo-continuum-normalized spectrum. We see wide wing-like features, most prominently here at the edges of the blue and green detectors. 
With renormalization, we bring the median flux of the spectrum closer to 1 (as shown in orange).

\subsection{Other data used in this paper} 
\label{xdata}
Additional data adopted in this paper includes stellar ages, used for calculating $\log{g}$ offsets between the isochrones and the APOGEE stars (Section~\ref{offsets}).  We use ages ('AGE\_LOWESS') from the astroNN Value Added Catalog \citep{Leung2019,Mackereth2019}\footnote{\url{https://www.sdss.org/dr16/data_access/value-added-catalogs/?vac_id=the-astronn-catalog-of-abundances,-distances,-and-ages-for-apogee-dr16-stars}}, derived using a neural network trained on the stellar spectra from DR14~\citep{jonsson2018}. 

We also use a set of M31 GC spectra, which were observed as part of an APOGEE ancillary project \citep{Zasowski2013} as well as a custom analysis of these data performed by \citet{Sakari2016}, for validation of our models (Section~\ref{gc}).
These spectra were made using the APOGEE \texttt{apVisit} files, with an approach nearly identical to the automated APOGEE pipeline that produces the combined \texttt{apStar} files\footnote{\url{https://www.sdss.org/dr16/irspec/spectral_combination/}}.  The difference is that we did not use the measured visit-level RVs when combining the visit spectra, because we found that some of these RVs were discrepant from the cluster mean; our combined GC spectra are thus in the observed frame, not the stellar rest frame.

\section{Generating Simple Stellar Population Models}
\label{generation}
In this section, we describe the steps in making a SSP spectral model along with how we use each ingredient. We adopt the Isochrone Synthesis method based on equation (1) from Section 2 of \citet{conroy2013}:

\begin{equation}
f_{\rm SSP}(t,Z) = \int_{m_{up}(t)}^{m_{lo}}f_{star} [T_{\rm eff}(M), \log{g} (M)|t,Z]\phi(M)dM,
\end{equation}

where $f_{\mathrm{star}}$ is the spectrum of a star with an initial stellar mass $M$. The isochrones with an initial mass function $\phi$(M) relate the T$_\mathrm{eff}$--$\log{g}$ distributions to an age ($t$) and chemical composition (Z). The model spectrum $f_{\mathrm{SSP}}$ represents the integration (from a lower limit $m_{lo}$, usually considered to be the hydrogen burning limit, to an upper limit $m_{up}(t)$ imposed by stellar evolution) of all the individual spectra from APOGEE along the isochrone, weighted by the fractional luminosity at a given T$_\mathrm{eff}$ and $\log{g}$ from the isochrone in the HR diagram. 

The isochrones for our SSP generation span a range in age, [M/H], and [$\alpha$/M] as described in Section~\ref{description}.

\subsection{Generating Synthetic Stellar Populations} 
\label{genssp}

Using an isochrone at a given [M/H] and [$\alpha$/M], we generate an SSP using the \citet{Kroupa2001} IMF with an initial mass of $10^{6}$ $\textup{M}_\odot$. Based on tests we ran using different initial masses \citep[e.g.,][]{Beerman2012} from $10^{5}$ $\textup{M}_\odot$ -- $10^{8}$ $\textup{M}_\odot$, $10^{6}$ $\textup{M}_\odot$  was selected as the minimum one needed to ensure that multiple SSPs generated from the same isochrone did not differ in the total number of stars or total luminosity by more than 1\%, due to stochastic sampling.

Our synthetic SSPs are generated using a package called PyPopStar \citep{Hosek2020}. This is a python package that generates synthetic single-age, single-metallicity population.  
We generate the synthetic SSP using our isochrones (age, metallicity and mass) and the initial mass of $10^{6}$ $\textup{M}_\odot$. The conditions to generate the SSP are:
\begin{itemize}
    \item Isochrone object: Isochrones from Padova or MIST (Section \ref{isochrones}).
    \item Initial mass = $10^{6}$~M$_\odot$
    \item IMF = \citet{Kroupa2001}
    \item Stellar Multiplicity: Defines the properties of multiple systems in a population based on the IMF. This function defines how the stellar masses influence the synthetic population. We use the 'MultiplicityUnresolved()' with parameters defined in \citet{Lu2013}.
    \item Synthetic star clusters: Using the isochrone object, IMF and the initial mass, we use the resolved star cluster function 'ResolvedCluster()' which interpolates the isochrone points within the mass limits to generate the synthetic population.
\end{itemize}

Each synthetic SSP generated is characterized by an age, [M/H], and [$\alpha$/M] and describes the T$_\mathrm{eff}$, $\log{g}$, and luminosity of stars at a given mass of $10^{6}$~M$_\odot$.

\subsection{Parameter Grid for SSP Spectral Model Generation} \label{paramgrid}
To construct our SSP spectral models (referred to as A-LIST models from here on), we need to obtain spectra that represent the simulated stars in a 4D parameter space (T$_\mathrm{eff}$, $\log{g}$, [M/H], and [$\alpha$/M]) for each age, [M/H], and [$\alpha$/M] isochrone. For this, we define 4D bins that span the range of T$_\mathrm{eff}$, $\log{g}$, [M/H], and [$\alpha$/M] needed for the simulated stars. We define bin sizes such that the stars in each bin have on average $<$1\% difference between their normalized spectra and the bin’s median spectrum:

\begin{enumerate}
\item Effective temperature (T$_\mathrm{eff}$): The full width of the bin $\Delta$T$_\mathrm{eff}$ = 50~K for 3000~K $\leq$ T$_\mathrm{eff}$ $\leq$ 5000~K and the full width of the bin $\Delta$T$_\mathrm{eff}$ = 150~K for 5000~K $\leq$T$_\mathrm{eff}$ $\leq$ 10000~K. We use a larger bin size for the hotter stars to increase the number of stars per bin (and the number of occupied bins), after confirming that the spectral variations within the bin are still within 1\%.

\item Surface gravity ($\log{g}$): $\Delta\log{g} = 0.2$~dex for $-1 < \log{g} < 5$. 

\item Metallicity ([M/H]): $\rm \Delta[M/H] = 0.3$~dex for $\rm -2.3 \leq [M/H] < -1.1$ and $\rm \Delta[M/H] = 0.2$~dex for $\rm -1.1 \leq [M/H] \leq +0.5$.

\item $\alpha$-abundance ([$\alpha$/M]): We define our [$\alpha$/M] bins based on the T$_\mathrm{eff}$ of the stars. For all stars having T$_\mathrm{eff}$ $\leq$ 5700K and $\rm -0.25 \leq [\alpha/M] < +0.45$, ~$\rm \Delta[\alpha/M] = 0.1$~dex. For hotter temperatures T$_\mathrm{eff}$ $>$ 5700K (i.e., the main sequence stars), there are insufficient $\alpha$--enhanced stars in our sample. However, because these hot stars have weaker lines, we found that the spectral variations at a given [M/H], log~g and T$_\mathrm{eff}$ bin across all $[\alpha/M]$ are still within 1\% at these temperatures.  Therefore, for these hot temperature bins we select all stars regardless of $\alpha$--enhancement in a given [M/H], T$_\mathrm{eff}$ and log~g bin.
\end{enumerate}

\begin{figure}[!htbp]
\includegraphics[width=\linewidth]{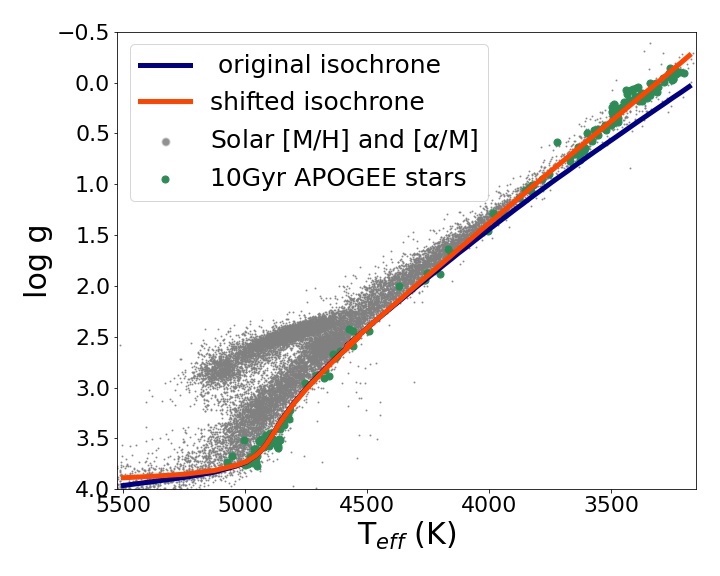}
\caption{Example of shifting an isochrone to account for the $\log{g}$ offset (Section~\ref{offsets}). The isochrone used here has an age of 10 Gyr, solar [M/H], and solar [$\alpha$/M]. The original isochrone is shown in blue. The isochrone is shifted by a function defined by the offset between the stars (solar [M/H], solar [$\alpha$/M] shown in grey) of the appropriate age (big green points) and the isochrone points. The shifted isochrone is shown in orange.}
\label{logoffset}
\end{figure}

\subsection{Surface Gravity Offset} 
\label{offsets}
When matching both the Padova and MIST isochrones with the empirical stellar library, we find that there are offsets in surface gravity and temperature along the upper part of the RGB.  Similar offsets along the upper RGB have been seen in previous work \citep[e.g.,][]{Serenelli2017, Durbin2020}. The uncertainties in the APOGEE $\log{g}$ are too small to account for these offsets, and the size of the offset varies for the different isochrone sets. Hence, to account for this difference, we shift the isochrone by a function defined by the offset between the stars and the isochrone in the RGB phase of evolution.

For each isochrone corresponding to an age, [M/H] and [$\alpha$/M], we select the APOGEE stars having the same [M/H] and [$\alpha$/M]. Based on the astroNN ages for APOGEE (Section~\ref{xdata}), we further select stars that are the closest 10\% in age to the isochrone. We then calculate the $\log{g}$ offset in the RGB evolutionary phase of the isochrone ($3000~K \leq T_{\rm eff} \leq 5500~K$) from the mean $\log{g}$ of the APOGEE stars in each temperature bin (Section~\ref{paramgrid}). A 3rd order polynomial is fit to this offset and the isochrones are shifted to match the APOGEE stars.

Figure~\ref{logoffset} shows an example of this process. A 10 Gyr, solar [M/H], solar [$\alpha$/M] isochrone (blue line) is plotted over the solar [M/H], solar [$\alpha$/M] APOGEE stars ($\sim$39,000 stars; shown as small grey points). The 10\% of APOGEE stars with astroNN ages closest to 10~Gyr in the RGB evolutionary phase are  shown as big green points ($\sim$170 stars). The isochrone shifted in $\log{g}$ to match these points is shown as an orange line. For the shift shown in Figure~\ref{logoffset}, the standard deviation of the difference between the shifted and unshifted spectra is just 0.14\%, while the maximum difference in the spectra is $\sim$1\%.

\subsection{Incorporating the APOGEE Spectral Library} 
\label{spectralmodel}
To obtain the integrated-light model spectrum of an SSP, the synthetic SSPs generated from the isochrones and the APOGEE stars are both binned in the 4D cube (Section~\ref{paramgrid}). Each bin is assigned a weight based on the fractional luminosity it contains, from the total synthetic SSP. In addition, each bin is assigned a median spectrum calculated from all of the APOGEE stars in that bin. Then, the final A-LIST model SSP is obtained by summing the median spectra from all the bins, each scaled by its luminosity weight determined from the binned synthetic SSP.

\section{Description of A-LIST Models} 
\label{description}
In this section, we describe the integrated near-infrared SSP model spectra created from APOGEE and synthetic SSPs. A-LIST contains SSP spectral models based on Padova isochrones and MIST isochrones, available at a range of ages, metallicities, and $\alpha$--abundances:
\begin{enumerate}
\item Age (Gyr): 2.0, 3.0, 4.0, 5.0, 6.0, 7.0, 8.0, 9.0, 10.0, 11.0, 12.0
\item Metallicity ([M/H]): -2.2, -1.9, -1.6, -1.3, -1.0, -0.8, -0.6, -0.4, -0.2, 0.0, +0.2, +0.4
\item $\alpha$--abundances ([$\alpha$/M]): -0.2, -0.1, 0.0, +0.1, +0.2, +0.3, +0.4
\end{enumerate}

\begin{figure*}
\includegraphics[width=\textwidth]{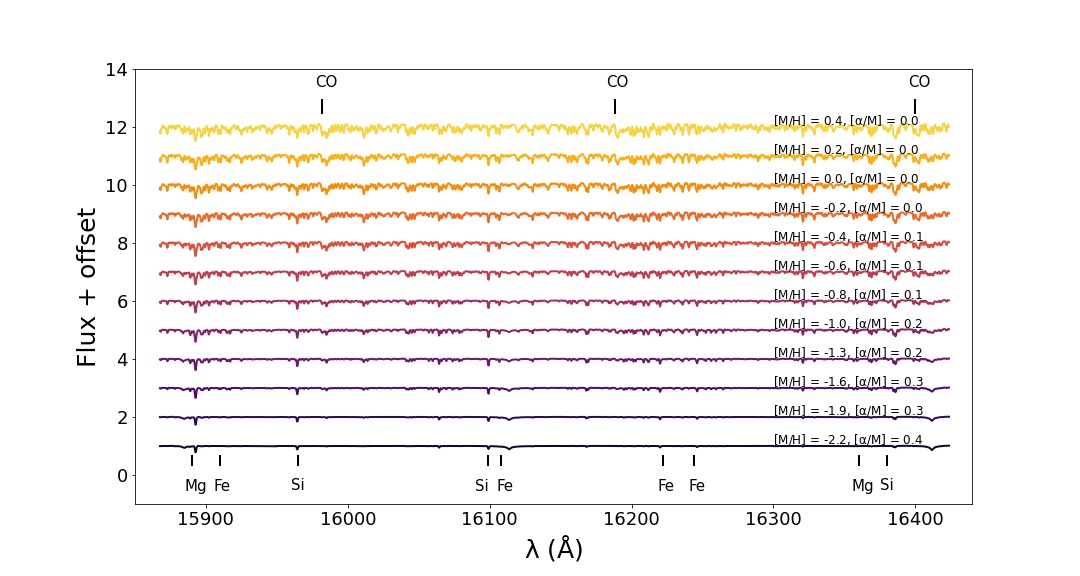}
\caption{Model spectra generated for 10~Gyr Padova--based SSPs ranging in [M/H] and [$\alpha$/M]. Shown here is a small wavelength range ($15850-16440$~\AA). For each [M/H], the model with the highest recovered fractional luminosity (Section~\ref{lumfrac}, Figure~\ref{validplot}(a)) is shown. Some Mg, Si, Fe lines are highlighted for this wavelength range along with the CO bandheads.}
\label{sspspectra}
\end{figure*}

\begin{figure*}[htbp!]
\centering
\gridline{\fig{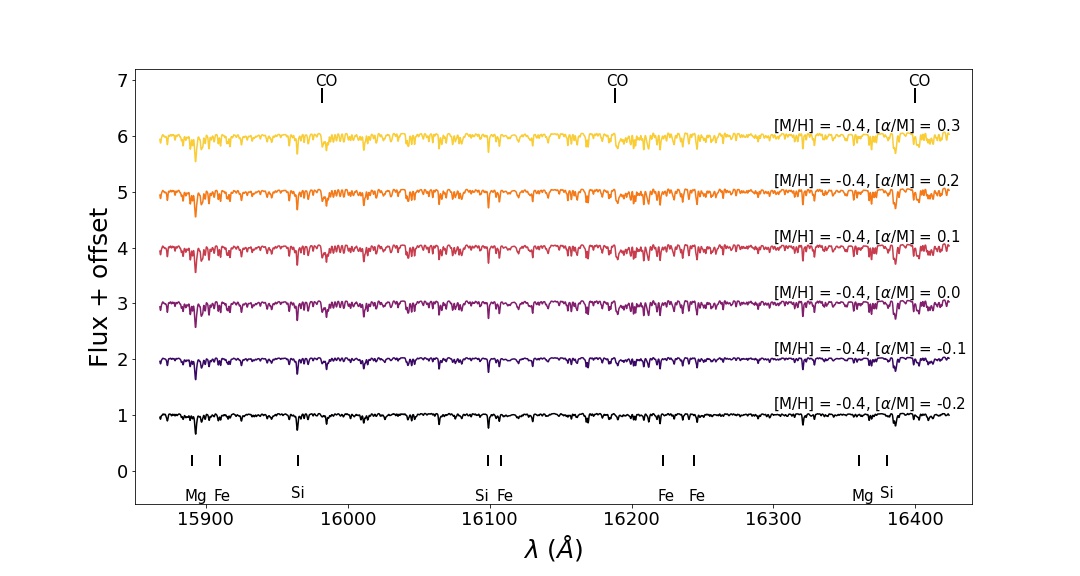}{0.67\textwidth }{(a)}
\vspace{-8pt}
\fig{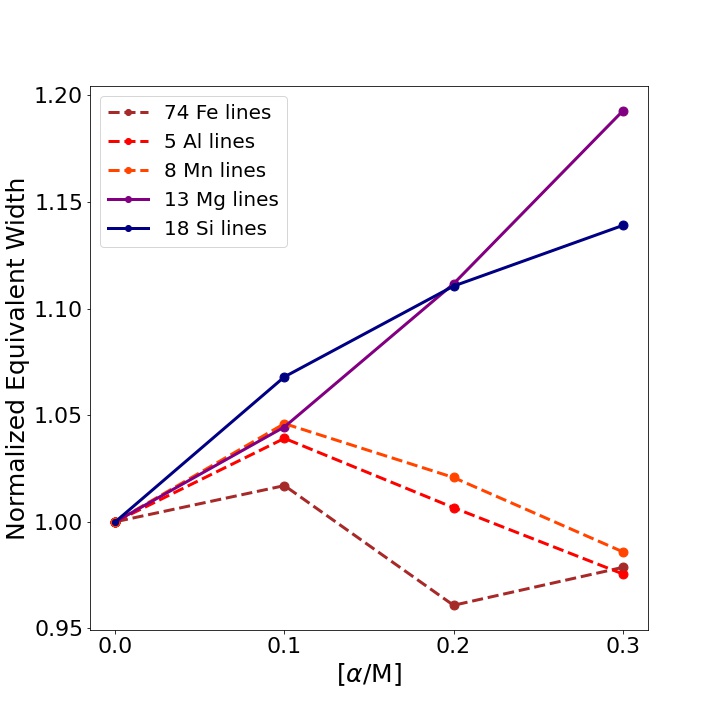}{0.36\textwidth}{(b)}}
\caption{[$\alpha$/M] variations in the models. Panel (a) shows the model spectra generated for 10~Gyr Padova-based SSPs with $\rm [M/H]=-0.4$, varying in [$\alpha$/M]. Shown here is a small wavelength range ($15850-16440$~\AA) with some of the major $\alpha$-bearing molecular bands and atomic lines labeled. We see the absorption lines get deeper from $\alpha$-poor to $\alpha$-rich models. Some of the Mg, Si and Fe lines used in panel (b) are highlighted, along with the CO bandheads. Panel (b) shows the variations in line equivalent widths (solid: $\alpha$ lines and dashed: non-$\alpha$ lines) for solar-$\alpha$ and $\alpha$-enhanced models. The circles are the
means of all the EWs of the respective element, normalized by the mean EW of the same element at solar-$\alpha$.}

\label{sspspectraalpha}
\end{figure*}

Figure~\ref{sspspectra} shows examples of 10~Gyr~old, Padova--based A-LIST spectral models varying in [M/H] and [$\alpha$/M] across the full range of [M/H].  The abundance change in the models is reflected in many of the absorption lines, strengthening  from metal poor to metal rich, e.g., in the Fe lines visible at  $\sim$15900~\AA. Figure~\ref{sspspectraalpha}(a) shows examples of 10~Gyr~old, Padova-based A-LIST spectral models with $\rm [M/H] = -0.4$ and different [$\alpha$/M]. An example of how models change with [$\alpha$/M] can be seen at the CO bandheads at 15990 \& 16180~\AA, as well as in the Mg line at 16365~\AA. In all cases these $\alpha$-element lines increase in strength from $\alpha$-poor to $\alpha$-rich models.

Figure \ref{sspspectraalpha}(b) shows the mean equivalent width (EW) of lines of different elements for each of the high fractional luminosity A-LIST models ($>$0.5; see Section \ref{lumfrac}) shown in Figure~\ref{sspspectraalpha}(a). We calculate the EW of non-blended absorption lines in the model spectra corresponding to Fe, Al, and Mn (non-$\alpha$ elements) and to Mg and Si ($\alpha$-elements). Each of the points shown in Figure~\ref{sspspectraalpha}(b) is the mean value of all the individual elemental EWs normalized to their element's [$\alpha$/M]=0 values.  
The total number of lines represented in this plot are 74 for Fe, 8 for Mn, 5 for Al, 13 for Mg, and 18 for Si. We see an increase in the $\alpha$-element EWs (solid lines), in contrast to the non-$\alpha$ element EWs (dashed lines). Indeed, the non-$\alpha$ EWs decrease at higher [$\alpha$/M], as expected with scaled-solar [M/H] held constant.

For each model, we also provide a spectrum that characterizes the standard deviation of the individual stellar spectra in each 4D bin used to create that model. The standard deviation of the spectra are measured at each spectral pixel in each 4D bin, and then these individual bin standard deviation spectra are added together using the same weights as the model spectra. 
This ``variance'' spectrum has a typical value of 1\% across all ages, [M/H], and [$\alpha$/M]. However, as we show in the next section, the dominant uncertainties in our model spectra are due to a lack of APOGEE stellar spectra for some models in log g--T$_\mathrm{eff}$ bins, which are not reflected in the provided variance spectrum.

\subsection{ Metrics for Assessing the Accuracy of A-LIST Models} \label{quality}

One simple method to validate the A-LIST models is to compare the differences between the models and an ideal synthetic stellar population model. In this section, we assume the synthetic SSPs to fully represent a stellar population of a given age, [M/H], and [$\alpha$/M]. The differences observed are due to either a lack of available stars in our empirical stellar library or our binning methods. We discuss the similarities between A-LIST models and the synthetic SSPs in terms of the luminosity recovered~(Section~\ref{lumfrac}) and the average T$_\mathrm{eff}$~(Section~\ref{avgtemp}). We also discuss the variation of the average metallicity and $\alpha$-abundance of A-LIST models from the defined bin center (Section~\ref{metvariation}).

\subsubsection{Fraction of the Synthetic SSP Luminosity Represented in A-LIST Models} \label{lumfrac}

\begin{figure}
\includegraphics[trim=5cm 0cm 4cm 2cm, clip, width=0.52\textwidth]{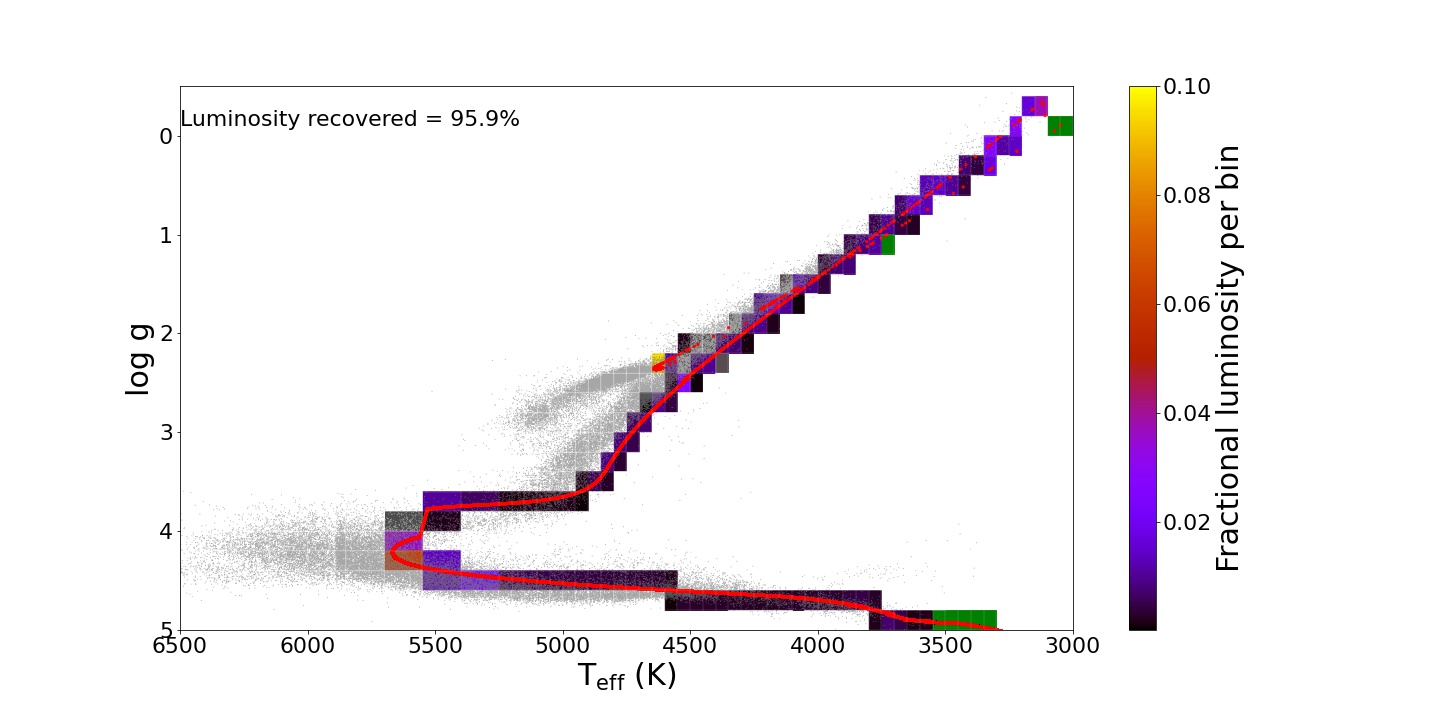}
\caption{HR diagram of a 10 Gyr, solar [M/H], solar [$\alpha$/M] SSP, colored by fraction of the total synthetic population luminosity in each bin. The APOGEE stars for this [M/H] and [$\alpha$/M] bin are shown as gray points. The SSP based on a Padova isochrone (for the given age, [M/H] and [$\alpha$/M]) is shown as red points.}
\label{hrdlumfrac}
\end{figure}

\begin{figure*}[htbp!]
\centering
\hspace{25pt}
\gridline{\fig{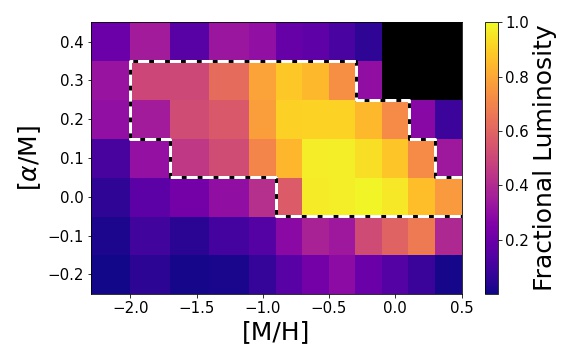}{0.34\linewidth}{(a)}
\fig{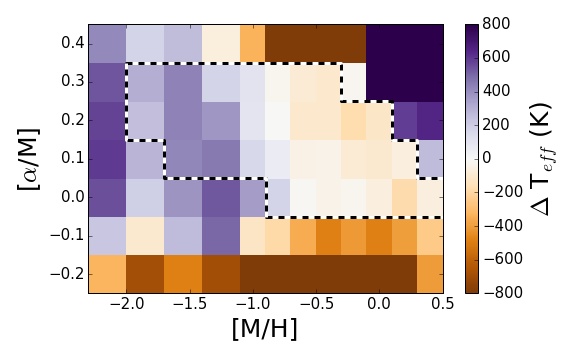}{0.34\linewidth}{(b)}
\fig{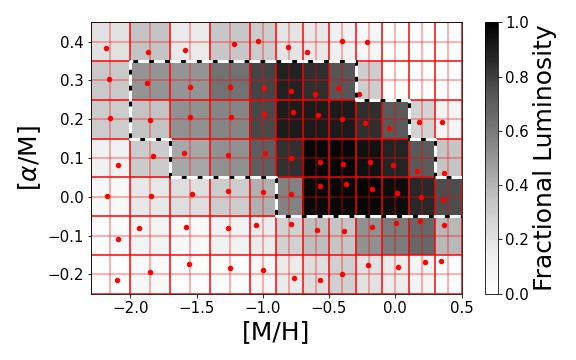}{0.34\linewidth}{(c)}}
\caption{Reliability of 10 Gyr Padova-based A-LIST models using the Padova-based synthetic SSPs as a function of [M/H] and [$\alpha$/M]. The bins that lie within the black and white dashed lines in each panel are the models that we recommend to use (Section~\ref{chi2test}). Panel~(a) shows the fractional luminosity of the A-LIST models generated using 10 Gyr synthetic SSPs (Section~\ref{lumfrac}). 
Panel~(b) shows the difference in the luminosity-weighted average temperature between 10 Gyr Padova-based synthetic SSPs and their corresponding A-LIST models (Section~\ref{avgtemp}). 
Panel~(c) shows the fractional luminosity of each A-LIST model (as in Panel~(a)) with the mean metallicity and $\alpha$-abundance values of each A-LIST model (Section~\ref{metvariation}). The red dots indicate the luminosity-weighted average values in each bin, and the red lines indicate the bin centers. These values are calculated excluding stars with  T$_\mathrm{eff}$ $>$5700~K (Section~\ref{paramgrid}).}
\label{validplot}
\end{figure*}
 
One important aspect of the A-LIST spectral models is how much of the total synthetic SSP luminosity is represented. 
Figure~\ref{hrdlumfrac} shows a HR Diagram that describes how we compute the fractional luminosity from the synthetic SSP for one A-LIST model (10~Gyr, solar [M/H], solar [$\alpha$/M]). The APOGEE stars with the same [M/H] and [$\alpha$/M] are shown as black points. The synthetic SSP is shown as red points. From this synthetic SSP we calculate the fraction of light in each of our spectral bins (Section~\ref{paramgrid}), which is indicated by the color of the bin. Bins that have a non-zero fraction of luminosity but no APOGEE stars in them --- i.e., bins that are not represented in the final A-LIST model spectrum --- are colored green. There are $\approx$8000 stars at solar [M/H] and solar [$\alpha$/M] (Figure~\ref{apogeeparam}(b)), enabling the high fractional luminosity here ($\sim$0.97).

Figure \ref{validplot}(a) shows the fractional luminosity of each 10~Gyr A-LIST spectral model as a function of [M/H] and [$\alpha$/M].  In general, the highest fractional luminosities coincide with the most-populated parts of chemical abundance space in the APOGEE sample.  In Section~\ref{chi2test} we discuss in detail how the fractional luminosity captured by the A-LIST models impacts the model properties and fits. We demonstrate there that the models with luminosity fraction below 0.32 are less reliable, and thus in Figure~\ref{validplot}(a) we highlight the reliable models with a white dashed outline.

\subsubsection{ Average temperature of A-LIST Models} 
\label{avgtemp}
Figure \ref{validplot}(b) shows the difference in the luminosity-weighted average T$_\mathrm{eff}$ for 10 Gyr A-LIST models, as a function of [M/H] and [$\alpha$/M], between the Padova-based synthetic SSPs and corresponding A-LIST models. The large negative differences are due to a lack of cool ($\sim$3500~K), low surface gravity stars in our APOGEE sample at particular metallicity and $\alpha$ combinations (e.g., at $\rm [M/H]=0.0$ and $\rm [\alpha/M]=-0.1$, where $\Delta T_\mathrm{eff} \sim -400$~K). The large positive differences are due to a lack of hot ($\sim$7000~K) main sequence stars; these are mostly in our lower metallicity models 
(e.g., at $\rm [M/H]=-1.3$ and $\rm [\alpha/M]=0.1$, where $\Delta T_\mathrm{eff} \sim 300$~K). In Section~\ref{chi2test} we assess how the change in luminosity-weighted average T$_\mathrm{eff}$ impacts the model properties.

\subsubsection{Variation in Metallicity and $\alpha$-abundance} \label{metvariation}
In Section \ref{paramgrid} we define the [M/H] and [$\alpha$/M] bins in our hypercube based on the desired model grid values. These bin values are assigned to the generated spectral models, assuming that the mean [M/H] and [$\alpha$/M] values of stars in the bin is the same as that of the bin center (e.g., in the [M/H] bin with bin-edges $\rm -1.1 \leq [M/H] < -0.9$, we assume the mean $\rm [M/H] = -1.0$). 

However, due to the non-uniform distribution of APOGEE stars across our bins, our mean [M/H] and [$\alpha$/M] are sometimes offset from the bin centers as shown in Figure \ref{validplot}(c).
The red points indicate the mean [M/H] and [$\alpha$/M] values for 10 Gyr Padova-based A-LIST models. These values are obtained by calculating the average [M/H] and [$\alpha$/M] weighted by the fractional luminosity of all the APOGEE stars with $T_\mathrm{eff}< \sim5700$~K (i.e., all stars that are binned by [$\alpha$/M]). The bin centers are shown by the inner red lines. 
For example, in the $\rm [\alpha/M] = -0.1$ and $\rm [M/H] = -1.0$ bin, we select stars with -1.1 $\leq$ [M/H] $<$ -0.9. Although the mean $\rm [M/H] = -1.0$ based on the bin edges, the actual mean $\rm [M/H] = -1.07$, based on the stars available in the bin.
The average absolute offset from the input [M/H] is 0.013 for all models with luminosity fraction above 0.32 (including those inside the dashed line in Figure \ref{validplot}(c)), while for [$\alpha$/M] this average absolute offset is 0.018.

\subsection{Difference Between Padova and MIST Based Models} 
\label{padmist}
\begin{figure}
\includegraphics[trim=0cm 0cm 4cm 2cm, clip, width=0.48\textwidth]{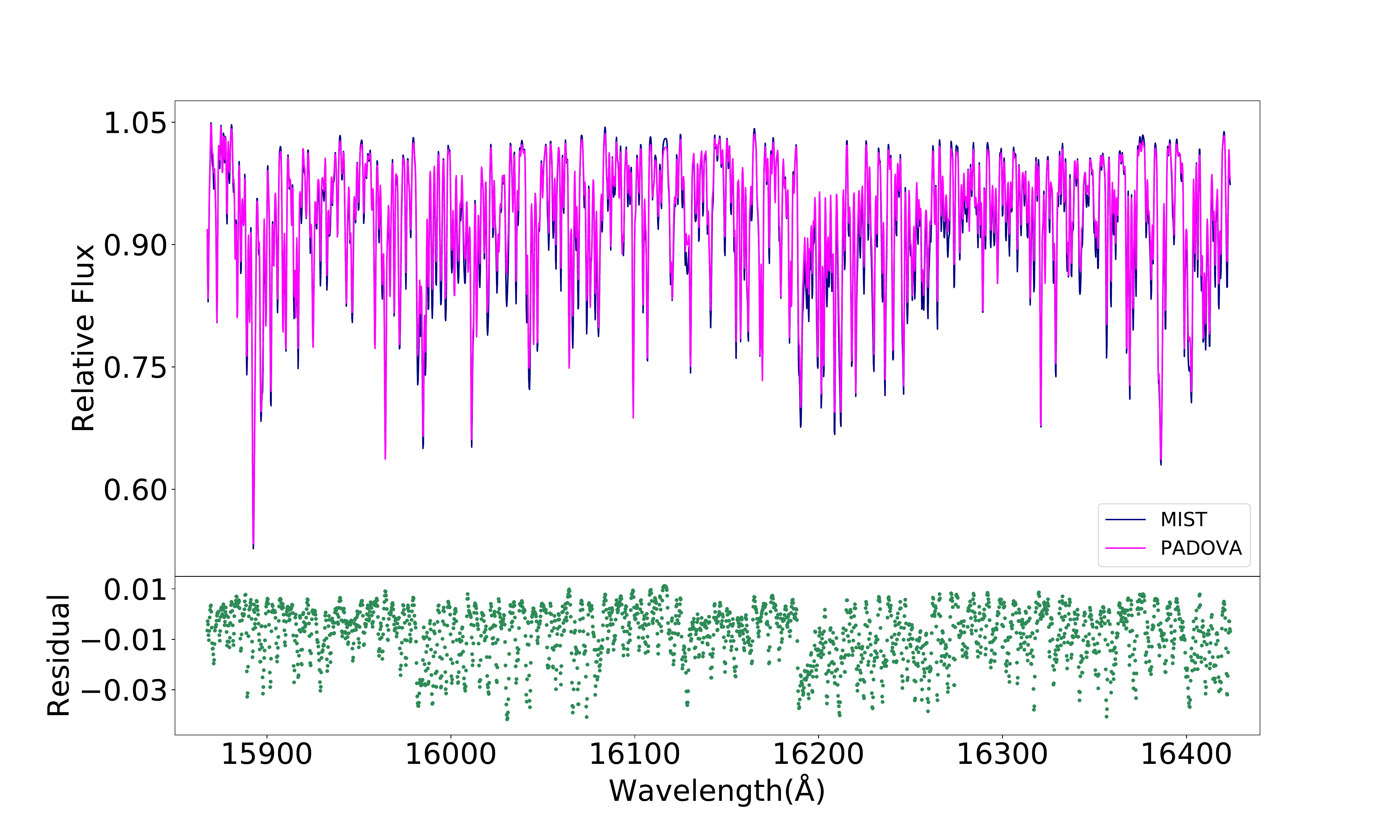}
\caption{Difference between a 10~Gyr, solar [M/H], solar [$\alpha$/M] Padova--based model (fuchsia) and MIST--based model (dark blue). Shown here are the model spectra for a small wavelength range ($15850-16440$~\AA). The residual (in green) is a simple difference between the two models.}
\label{padmistfig}
\end{figure}

A-LIST consists of spectral models generated using two different isochrone sets: Padova and MIST (Section \ref{isochrones}). Figure~\ref{padmistfig} shows a comparison between 10 Gyr, solar [M/H], solar [$\alpha$/M] MIST-based (shown in blue) and Padova-based (shown in fuchsia) A-LIST model spectra. The maximum difference is $\sim$3\%, and the standard deviation of the difference is $<$1\%.

The difference between Padova-based and MIST-based model spectra arises from the differences in the isochrones, primarily from the cooler temperature bins (briefly explained in Section \ref{isochrones}). For example, in Figure~\ref{padmistfig}, the TP-AGB stars in the MIST-based population have $T_\mathrm{eff} \sim 3200$~K, and in the Padova--based one, they have $T_\mathrm{eff} \sim 3100$~K. This $\sim$100~K difference leads to deeper absorption lines in the Padova models, with a maximum difference of 3\%.
In Section~\ref{gc} we show that while overall both models recover the metallicities and $\alpha$-abundances of globular clusters, these model differences lead to small differences in measurements for individual clusters.
We note that, consistent with the TP-AGB star difference discussed above, typically the Padova models have luminosity weighted effective temperatures that are $\sim$~100K hotter than the MIST models with the same population parameters.

\begin{figure}
\includegraphics[width=0.48\textwidth]{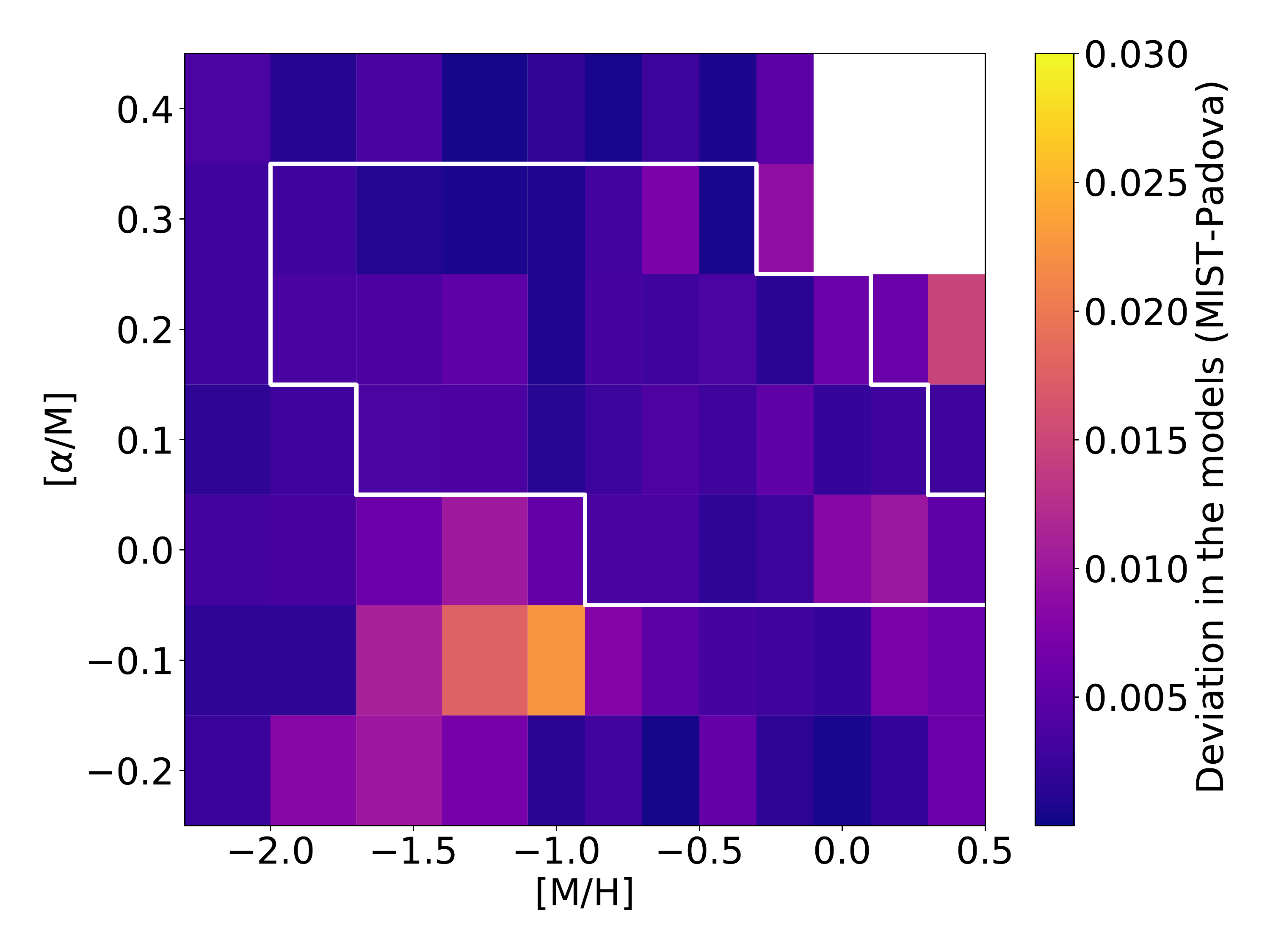}
\caption{The standard deviation of the difference between the 10~Gyr Padova- and MIST--based spectral models for each bin in [M/H] and [$\alpha$/M]. The white line is the same as the white dashed line shown in Figure~\ref{validplot}(a). A small difference is observed for the models of each age, [M/H], [$\alpha$/M].}
\label{padmistdiff}
\end{figure}

Figure \ref{padmistdiff}  shows the standard deviation in the difference between the 10 Gyr MIST--based and Padova--based A-LIST models models as a function of [M/H] and [$\alpha$/M]. We see that the difference in the models for the most populated parts of the chemical abundance space is $<$1.5\% of the normalized flux. The bins with larger deviations ($>$1.5\%) correspond to the least-populated parts of chemical abundance space of the APOGEE stars (like very metal-poor, solar-$\alpha$ bins and metal-rich, $\alpha$-rich bins), which also correspond to regions lying outside the white ``high reliability'' contour (Section~\ref{chi2test}).
\\
\section{SPECTRAL MODEL VALIDATION} 
\label{validation}
In this section, we validate A-LIST models by recovering the kinematics, [M/H], and [$\alpha$/M] of M31 GCs and by comparing our A-LIST spectral library to the lower resolution E-MILES \citep{vazdekis2016} library.

\subsection{Fitting procedure} 
\label{ppxf}
We use the Penalized Pixel-Fitting method \citep[pPXF;][]{Cappellari2009, Cappellari2017}, a full-spectrum fitting code that calculates stellar kinematics and stellar population parameters from absorption-line spectra using a maximum penalized likelihood approach. We compute velocities, dispersions, and stellar population parameters for the GCs using A-LIST as templates, and to compare the age and [M/H] of our high resolution models with a low resolution spectral library.

\subsection{M31 GCs} 
\label{gc}
GCs are not perfect SSPs \citep[e.g.,][]{Gratton2012}, but are very close to being so; hence, they are useful for validating our A-LIST models by recovering the properties of previously studied clusters observed by APOGEE (Section~\ref{xdata}). We select 32 GCs with 10 $\leq$ SNR $\leq$ 150 \citep[7 of which were studied in][]{Sakari2016}.

To determine the properties of the GCs, we fit each of our A-LIST models one-at-a-time to the GCs over the wavelength range of the green and red detectors (i.e., $15850-16960$~\AA). We do not use the blue detector due to the presence of high persistence in many spectra \citep[see Section 3.4 of][]{Majewski2017}. We take the likelihood-weighted average for age, [M/H], and [$\alpha$/M] from the three fits with smallest reduced $\chi^2$ values. We use all A-LIST models with age $>$6~Gyr to reduce the time needed to run the code. The uncertainties on these fits are calculated in two different ways: for radial velocity and the velocity dispersion, the systematic uncertainty on the best fit model is determined by a simple bootstrapping for $n = 100$. For [M/H] and [$\alpha$/M], we calculate the uncertainty using the relative $\chi^2$ of the best fit model relative to other models. More specifically, we derive our uncertainties in [M/H] and [$\alpha$/M] by using a $\Delta \chi^2$ limit corresponding to a 1$\sigma$ uncertainty, deriving both lower and upper limits.

Table~\ref{tab:1} shows the radial velocity, velocity dispersion, [M/H], [$\alpha$/M], and age of the M31 GCs determined using the A-LIST library, both Padova--based and MIST--based, along with the uncertainties.

\begin{figure}[h!]
\includegraphics[trim=1cm 12cm 0cm 0cm, clip, width=0.5\textwidth]{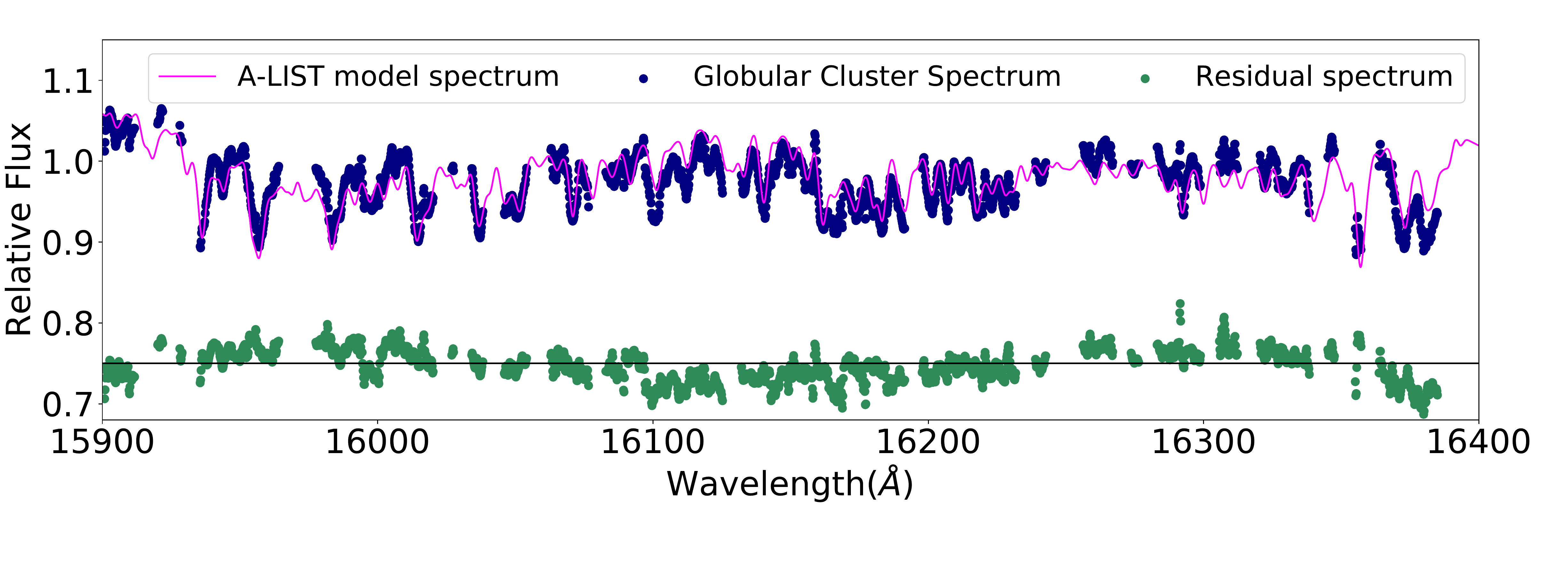}
\caption{Example of a best fit model to a high SNR M31 GC spectrum (B127-G185, $\rm SNR=150$). Here we are fitting a 10~Gyr, $\rm [M/H] = -0.6$, $\rm [\alpha/M] = 0.1$ A-LIST model spectrum.  We note that some regions of the spectra have been excluded from the fit due to the presence of strong sky lines; these sections have been excluded from the fit and are not plotted in the spectrum or residuals.}
\label{b127}
\end{figure}

Figure~\ref{b127} shows an example of a high SNR GC spectrum (B127-G185, SNR = 150) with its best fit A-LIST Padova--based model spectrum, for a small section of the wavelength range ($15850-16440$~\AA). The GC spectrum is first masked for bad sky lines as well as in regions where the error on the flux is high. The residuals are very small (standard deviation $\sim0.013\%$)

\begin{figure*}[htbp!]
\centering
\includegraphics[width=0.9\textwidth]{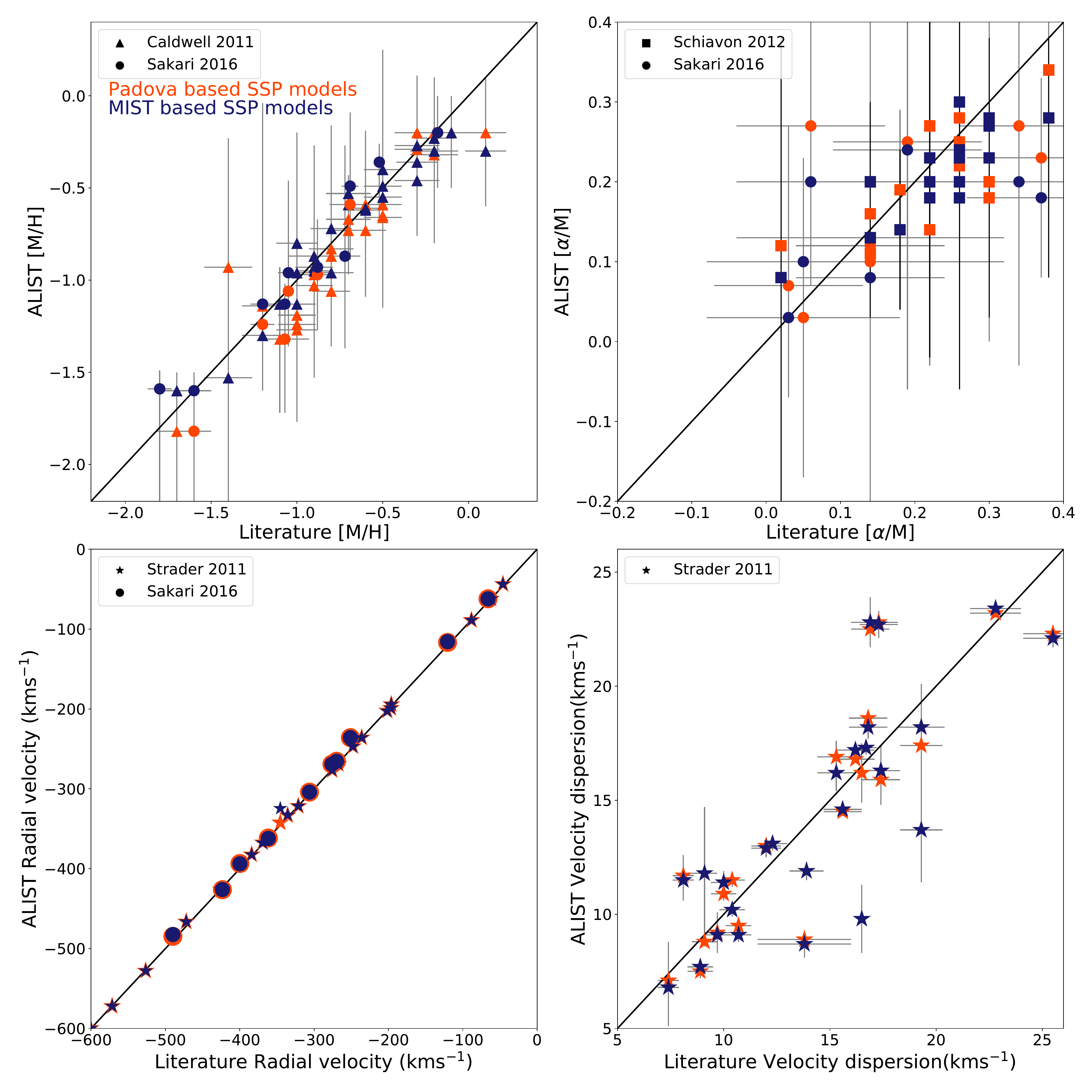}
\caption{Properties of M31 GCs recovered using A-LIST spectral models. The orange and blue symbols denote values from Padova- and MIST--based A-LIST models, respectively. The symbol shapes indicate the literature source: circles for \citet{Sakari2016}, triangles for \citet{Caldwell2011}, stars for \citet{Strader2011}, and squares for \citet{Schiavon2012}.}
\label{m31prop}
\end{figure*}

Figure~\ref{m31prop} compares our best-fit values from the A-LIST models with literature values. Due to a lack of available [$\alpha$/M] literature measurements for our M31 GCs, we use [Mg/Fe] derived from the line indices in \citet{Schiavon2012} to compare with A-LIST [$\alpha$/M]. The literature comparison for the other parameters are obtained from \citet{Strader2011} for velocity dispersion and radial velocities and from \citet{Caldwell2011} for [Fe/H]. We also include radial velocities, metallicity, and [$\alpha$/Fe] estimates for these GCs from \citet{Sakari2016}.

The dynamical and chemical properties recovered from the A-LIST models agree well with these previous literature values (Table~\ref{tab:2}). For Padova-based (MIST-based values in parentheses) we find a median offset (literature -- A-LIST) of 0.09 (0.04) in [M/H] and  0.02 (0.02) in [$\alpha$/M], while the median absolute deviation is 0.1 (0.08) in [M/H] and 0.04 (0.04) in [$\alpha$/M]. These median absolute deviations are comparable to or smaller than the size of the literature errors on these quantities.

Overall, we find that the MIST-based model metallicities match more closely those of \citet{Caldwell2011}, while the Padova-based models give somewhat lower metallicities.  This lower metallicity is consistent with the deeper lines seen for the same population parameters in the Padova-based templates relative to the MIST-based templates, due to the cooler AGB stars and luminosity-weighted $T_\mathrm{eff}$. 
A Spearman rank correlation between our A-LIST [$\alpha$/M] values and the [Mg/Fe] of \citet{Schiavon2012} gives a correlation coefficient of 0.64 and a p-value of 1.37x10$^{-4}$, suggesting a very significant correlation.

\begin{deluxetable*}{cccccccccccc}
\tabletypesize{\scriptsize}
\tablewidth{-25pt}
\tablecolumns{12}
\tablecaption{\label{tab:1} Table containing the age, [M/H], [$\alpha$/M], RV, and $\sigma$  retrieved from fitting the Padova--based and MIST--based A-LIST spectral models to 32 selected M31 GCs.}
\tablehead{\\\nocolhead{} & \nocolhead{} & \multicolumn{5}{c}{MIST--based SSP} &  \multicolumn{5}{c}{Padova--based SSP} \\
\hline
\colhead{Globular cluster} & \colhead{SNR} & \colhead{RV} & \colhead{$\sigma$} & \colhead{[M/H]} & \colhead{[$\alpha$/M]} & \colhead{Age} & \colhead{RV} & \colhead{$\sigma$} & \colhead{[M/H]} & \colhead{[$\alpha$/M]} & \colhead{Age} \\
\nocolhead{} & \nocolhead{} & \colhead{(km~s$^{-1}$)} & \colhead{(km~s$^{-1}$)} & \nocolhead{} & \nocolhead{} & \colhead{(Gyr)} & \colhead{(km~s$^{-1}$)} & \colhead{(km~s$^{-1}$)} & \nocolhead{} & \nocolhead{} & \colhead{(Gyr)}}
\startdata
B006-G058 & 25 & -235.9$\pm$0.3 & 12.9$\pm$0.4 & -0.5$^{+0.3}_{-0.4}$ & 0.2$^{+0.2}_{-0.1}$ & 10.3$^{+1.7}_{-2.1}$ & -235.9$\pm$0.3 & 13.0$\pm$0.2 & -0.6$^{+0.3}_{-0.3}$& 0.3$^{+0.1}_{-0.1}$& 10.3$^{+1.1}_{-1.6}$ \\
B012-G064 & 23 & -362.1$\pm$ 1.9 & 13.7$\pm$2.3 & -1.6$^{+0.8}_{-0.1}$ & 0.1$^{+0.3}_{+0.3}$ & 10.3$^{+1.2}_{-1.5}$ & -361.6$\pm$1.9 & 17.4$\pm$2.7 & -1.8$^{+0.5}_{-0.2}$ & 0.1$^{+0.2}_{+0.1}$ & 10.3$^{+2.1}_{-1.2}$ \\
B023-G078 & 180 & -440.3$\pm$0.4 & 28.5$\pm$0.4 & -0.5$^{+0.3}_{-0.1}$ & 0.1$^{+0.2}_{+0.1}$ & 6.7$^{+2.6}_{-2.3}$ & -442.3$\pm$0.6 & 29.5$\pm$0.4 & -0.5$^{+0.3}_{-0.1}$ & 0.1$^{+0.1}_{-0.1}$ & 6.3$^{+3.5}_{-3.1}$ \\
B025-G084 & 10 & -202.7$\pm$2.6 & 11.8$\pm$2.9 & -1.5$^{+0.5}_{-0.1}$ & 0.1$^{+0.1}_{-0.1}$ & 8.3$^{+0.7}_{-2.2}$ & -202.2$\pm$0.8 & 8.8$\pm$0.2 & -0.9$^{+0.5}_{-0.3}$ & 0.2$^{+0.1}_{-0.1}$ & 8.4$^{+1.3}_{-1.2}$ \\
B045-G108 & 17 & -426.1$\pm$0.5 & 11.4$\pm$0.5 & -0.9$^{+0.2}_{-0.1}$ & 0.2$^{+0.1}_{-0.1}$ & 10.6$^{+2.4}_{-2.5}$ & -426.3$\pm$0.3 & 10.9$\pm$0.3 & -1.0$^{+0.3}_{-0.3}$ & 0.2$^{+0.2}_{-0.2}$ & 10.6$^{+2.3}_{-2.5}$ \\
B063-G124 & 34 & -304.1$\pm$0.4 & 18.2$\pm$0.5 & -1.0$^{+0.1}_{-0.5}$ & 0.2$^{+0.1}_{-0.1}$ & 10.3$^{+1.4}_{-2.6}$ & -304.1$\pm$0.3 & 18.6$\pm$0.3 & -1.0$^{+0.3}_{-0.3}$ & 0.2$^{+0.2}_{-0.2}$ & 8.3$^{+1.2}_{-1.5}$ \\
B068-G130 & 23 & -321.6$\pm$0.3 & 9.1$\pm$0.3 & -0.2$^{+0.3}_{-0.3}$ & 0.2$^{+0.1}_{-0.1}$ & 9.7$^{+1.2}_{-1.8}$ & -321.6$\pm$0.3 & 9.5$\pm$0.3 & -0.2$^{+0.1}_{-0.3}$ & 0.2$^{+0.1}_{-0.1}$ & 9.7$^{+2.6}_{-2.7}$ \\
B088-G150 & 32 & -482.4$\pm$1.5 & 9.8$\pm$1.5 & -1.6$^{+0.7}_{-0.1}$ & 0.0$^{+0.1}_{-0.1}$ & 8.5$^{+2.6}_{-0.8}$ & -484.7.0$\pm$1.4 & 16.2$\pm$1.3 & -1.6$^{+0.1}_{-0.3}$ & 0.1$^{+0.2}_{-0.1}$ & 8.8$^{+2.3}_{-1.2}$ \\
B103-G165 & 98 & -367.6$\pm$ 0.3 & 17.3$\pm$0.3 & -0.4$^{+0.3}_{-0.4}$ & 0.3$^{+0.1}_{-0.1}$ & 11.3$^{+0.9}_{-1.2}$ & -367.3$\pm$0.2 & 17.3$\pm$0.2 & -0.6$^{+0.4}_{-0.2}$ & 0.2$^{+0.2}_{-0.2}$ & 10.7$^{+0.5}_{-2.6}$ \\
B107-G169 & 47 & -332.8$\pm$0.8 & 16.3$\pm$1.1 & -1.0$^{+0.2}_{-0.2}$ & 0.1$^{+0.1}_{-0.1}$ & 11.3$^{+0.3}_{-2.1}$ & -333.2$\pm$0.8 & 15.9$\pm$1.1 & -1.2$^{+0.1}_{-0.1}$ & 0.2$^{+0.2}_{-0.1}$ & 11.3$^{+1.2}_{-1.3}$ \\
B109-G170 & 23 & -620.3$\pm$0.2 & 7.7$\pm$0.3 & -0.3$^{+0.5}_{-0.1}$ & 0.3$^{+0.1}_{-0.1}$ & 10.1$^{+2.1}_{-1.2}$ & -620.3$\pm$0.2 & 7.5$\pm$0.3 & -0.3$^{+0.3}_{-0.3}$ & 0.3$^{+0.1}_{-0.1}$ & 10.0$^{+0.8}_{-2.3}$ \\
B112-G174 & 20 & -276.5$\pm$0.2 & 10.2$\pm$0.3 & -0.5$^{+0.3}_{-0.1}$ & 0.2$^{+0.2}_{-0.2}$ & 9.0$^{+2.4}_{-3.4}$ & -276.7$\pm$0.2 & 11.5$\pm$0.3 & -0.2$^{+0.2}_{-0.2}$ & 0.1$^{+0.2}_{-0.1}$ & 10.3$^{+3.8}_{-0.8}$ \\
B115-G177 & 36 & -599.7$\pm$0.3 & 13.1$\pm$0.3 & -0.3$^{+0.3}_{-0.2}$ & 0.2$^{+0.1}_{-0.1}$ & 10.8$^{+1.7}_{-3.2}$ & -599.7$\pm$0.2 & 13.1$\pm$0.2 & -0.2$^{+0.3}_{-0.3}$ & 0.2$^{+0.1}_{-0.1}$ & 10.6$^{+2.3}_{-0.7}$ \\
B127-G185 & 151 & -527.8$\pm$0.4 & 23.4$\pm$0.4 & -0.6$^{+0.1}_{-0.1}$ & 0.2$^{+0.1}_{-0.1}$ & 10.1$^{+0.6}_{-0.8}$ & -527.8$\pm$0.2 & 23.2$\pm$0.3 & -0.7$^{+0.2}_{-0.2}$ & 0.2$^{+0.1}_{-0.1}$ & 10.3$^{+0.8}_{-2.3}$ \\
B128-G187 & 15 & -382.4$\pm$0.8 & 11.5$\pm$0.9 & -0.6$^{+0.1}_{-0.3}$ & 0.3$^{+0.1}_{-0.1}$ & 10.9$^{+0.3}_{-0.2}$ & -382.3$\pm$0.8 & 11.7$\pm$0.9 & -0.6$^{+0.4}_{-0.3}$ & 0.2$^{+0.1}_{-0.1}$ & 11.1$^{+0.9}_{-0.8}$ \\
B129 & 59 & -413.7$\pm$0.5 & 18.2$\pm$0.6 & -1.0$^{+0.2}_{-0.4}$ & 0.2$^{+0.1}_{-0.2}$ & 10.4$^{+2.5}_{-2.3}$ & -43.7$\pm$0.3 & 18.2$\pm$0.3 & -0.8$^{+0.1}_{-0.1}$ & 0.2$^{+0.1}_{-0.2}$ & 9.6$^{+3.7}_{-3.7}$ \\
B131-G189 & 105 & -466.4$\pm$0.6 & 22.7$\pm$0.6 & -0.5$^{+0.1}_{-0.1}$ & 0.2$^{+0.1}_{-0.1}$ & 9.3$^{+2.6}_{-1.8}$ & -466.3$\pm$0.4 & 22.8$\pm$0.6 & -0.7$^{+0.3}_{-0.2}$ & 0.2$^{+0.1}_{-0.2}$ & 9.1$^{+0.8}_{-2.3}$ \\
B151-G205 & 125 & -324.5$\pm$0.5 & 22.1$\pm$0.4 & -0.6$^{+0.2}_{-0.2}$ & 0.1$^{+0.1}_{-0.1}$ & 9.6$^{+2.7}_{-2.1}$ & -342.3$\pm$0.3 & 22.3$\pm$0.2 & -0.7$^{+0.1}_{-0.3}$ & 0.2$^{+0.1}_{-0.1}$ & 10.3$^{+1.0}_{-12}$ \\
B153 & 50 & -246.9$\pm$0.3 & 11.9$\pm$0.4 & -0.3$^{+0.2}_{-0.2}$ & 0.2$^{+0.1}_{-0.1}$ & 10.3$^{+2.5}_{-2.6}$ & -246.9$\pm$0.2 & 11.9$\pm$0.2 & -0.3$^{+0.2}_{-0.3}$ & 0.3$^{+0.1}_{-0.1}$ & 10.8$^{+2.1}_{-2.7}$ \\
B171-G222 & 72 & -269.1$\pm$0.3 & 14.6$\pm$0.3 & -0.4$^{+0.1}_{-0.2}$ & 0.2$^{+0.1}_{-0.1}$ & 10.1$^{+0.4}_{-0.6}$ & -269.1$\pm$0.2 & 14.5$\pm$0.2 & -0.4$^{+0.2}_{-0.3}$ & 0.2$^{+0.1}_{-0.1}$ & 9.8$^{+1.8}_{-0.5}$ \\
B180-G231 & 12 & -198.9$\pm$0.3 & 9.1$\pm$0.9 & -1.0$^{+0.4}_{-0.4}$ & 0.2$^{+0.1}_{-0.1}$ & 9.0$^{+3.2}_{-2.4}$ & -198.9$\pm$0.7 & 9.2$\pm$0.9 & -1.0$^{+0.5}_{-0.5}$ & 0.1$^{+0.1}_{-0.2}$ & 9.0$^{+1.2}_{-1.2}$ \\
B190-G241 & 10 & -88.8$\pm$0.3 & 6.8$\pm$1.7 & -1.3$^{+0.3}_{-0.5}$ & 0.2$^{+0.2}_{-0.1}$ & 10.6$^{+1.0}_{-2.1}$ & -88.3$\pm$0.4 & 7.1$\pm$1.8 & -1.1$^{+0.7}_{-0.3}$ & 0.1$^{+0.2}_{-0.1}$ & 10.4$^{+0.3}_{-1.4}$ \\
B193-G244 & 29 & -62.1$\pm$0.3 & 17.2$\pm$0.3 & -0.2$^{+0.3}_{-0.1}$ & 0.2$^{+0.2}_{-0.2}$ & 8.8$^{+2.4}_{-1.4}$ & -62.1$\pm$0.2 & 16.8$\pm$0.1 & -0.2$^{+0.2}_{-0.3}$ & 0.3$^{+0.2}_{-0.2}$ & 8.6$^{+4.6}_{-3.5}$ \\
B206-G257 & 58 & -194.3$\pm$0.6 & 16.2$\pm$0.7 & -1.1$^{+0.3}_{-0.1}$ & 0.2$^{+0.2}_{-0.1}$ & 9.7$^{+1.7}_{-1.2}$ & -194.2$\pm$0.7 & 16.9$\pm$0.1 & -1.1$^{+0.2}_{-0.2}$ & 0.2$^{+0.1}_{-0.1}$ & 9.7$^{+3.2}_{-1.7}$ \\
B213-G264 & 14 & -572.1$\pm$0.9 & 8.7$\pm$0.2 & -0.7$^{+0.3}_{-0.2}$ & 0.1$^{+0.1}_{-0.1}$ & 9.9$^{+0.9}_{-0.8}$ & -572.1$\pm$0.3 & 8.9$\pm$0.6 & -0.9$^{+0.3}_{-0.3}$ & 0.1$^{+0.1}_{-0.1}$ & 10.2$^{+1.5}_{-0.6}$ \\
B313-G036 & 19 & -420.3$\pm$3.2 & 10.9$\pm$3.9 & -0.8$^{+0.3}_{-0.3}$ & 0.1$^{+0.1}_{-0.2}$ & 10.0$^{+0.7}_{-1.2}$ & -426.2$\pm$0.8 & 7.8$\pm$1.7 & -1.2$^{+0.6}_{-0.3}$ & 0.1$^{+0.2}_{-0.1}$ & 11.00$^{+2.7}_{-0.8}$ \\
B373-G305 & 86 & -220.1$\pm$0.3 & 12.4$\pm$0.3 & -0.5$^{+0.1}_{-0.1}$ & 0.3$^{+0.2}_{-0.2}$ & 9.9$^{+2.3}_{-2.1}$ & -220.5$\pm$0.2 & 13.1$\pm$0.8 & -0.7$^{+0.3}_{-0.2}$ & 0.3$^{+0.1}_{-0.2}$ & 9.8$^{+3.1}_{-0.6}$ \\
B386-G322 & 48 & -393.8$\pm$0.5 & 12.8$\pm$0.9 & -1.1$^{+0.2}_{-0.1}$ & 0.2$^{+0.1}_{-0.2}$ & 9.3$^{+3.4}_{-1.3}$ & -393.5$\pm$0.9 & 14.3$\pm$1.5 & -1.3$^{+0.4}_{-0.1}$ & 0.3$^{+0.1}_{-0.2}$ & 9.3$^{+3.7}_{-2.6}$\\
B403-G348 & 40 & -265.6$\pm$0.5 & 9.9$\pm$0.6 & -0.9$^{+0.1}_{-0.2}$ & 0.1$^{+0.2}_{-0.1}$ & 10.4$^{+0.4}_{-1.2}$ & -265.5$\pm$0.3 & 9.7$\pm$0.4 & -0.9$^{+0.3}_{-0.1}$ & 0.0$^{+0.2}_{-0.2}$ & 10.2$^{+1.2}_{-1.5}$\\
B407-G352 & 49 & -296.2$\pm$0.5 & 9.2$\pm$0.4 & -0.6$^{+0.1}_{-0.8}$ & 0.1$^{+0.1}_{-0.1}$ & 11.4$^{+0.9}_{-0.9}$ & -296.7$\pm$0.2 & 9.7$\pm$0.9 & -0.7$^{+0.5}_{-0.3}$ & 0.2$^{+0.1}_{-0.1}$ & 11.4$^{+0.7}_{-1.8}$\\
B472-D064 & 55 & -115.8$\pm$0.5 & 22.8$\pm$1.6 & -1.1$^{+0.3}_{-0.1}$ & 0.1$^{+0.1}_{-0.1}$ & 9.5$^{+1.7}_{-1.6}$ & -116.7$\pm$0.3 & 22.5$\pm$0.2 & -1.2$^{+0.5}_{-0.2}$ & 0.1$^{+0.1}_{-0.1}$ & 9.5$^{+2.7}_{-2.2}$ \\
\enddata
\tablecomments{The selection of GCs (32 shown here from $\sim$185 available) is based on available literature values as well as based on physical pPXF fits with reasonable $\chi^2$ values. We eliminate those GCs that have a very noisy spectra and/or have a continuum issue in their chips.}
\end{deluxetable*}

\begin{deluxetable*}{c|cc|cc|cc|cc}
\tabletypesize{\scriptsize}
\tablewidth{-25pt}
\tablecolumns{9}
\tablecaption{\label{tab:2}Table showing the median offsets (literature -- A-LIST) and the median absolute deviation in A-LIST parameter values when fit to M31 GCs, compared to literature measurements (Figure~\ref{m31prop}).}
\tablehead{\\
&\multicolumn2c{RV}&\multicolumn2c{$\sigma$}&\multicolumn2c{[M/H]}&\multicolumn2c{[$\alpha$/M]}\\
&\multicolumn2c{(km~s$^{-1}$)}&\multicolumn2c{(km~s$^{-1}$)}&\multicolumn2c{(dex)}&\multicolumn2c{dex}}
\startdata
\hline
& MIST--based & Padova--based & MIST--based & Padova--based & MIST--based & Padova--based & MIST--based & Padova--based \\
\hline
median offset  &-1.5 & -1.3 & 0.4 & 0.2 & 0.04 & 0.09 & 0.02 & 0.02\\
\hline
median absolute deviation & 0.74 & 0.66 & 1.55 & 1.53 & 0.08 & 0.1 & 0.04 & 0.04\\
\hline
\enddata
\end{deluxetable*}

\begin{figure*}
\centering
\includegraphics[trim=1cm 1cm 0cm 0cm, clip, width=0.95\textwidth]{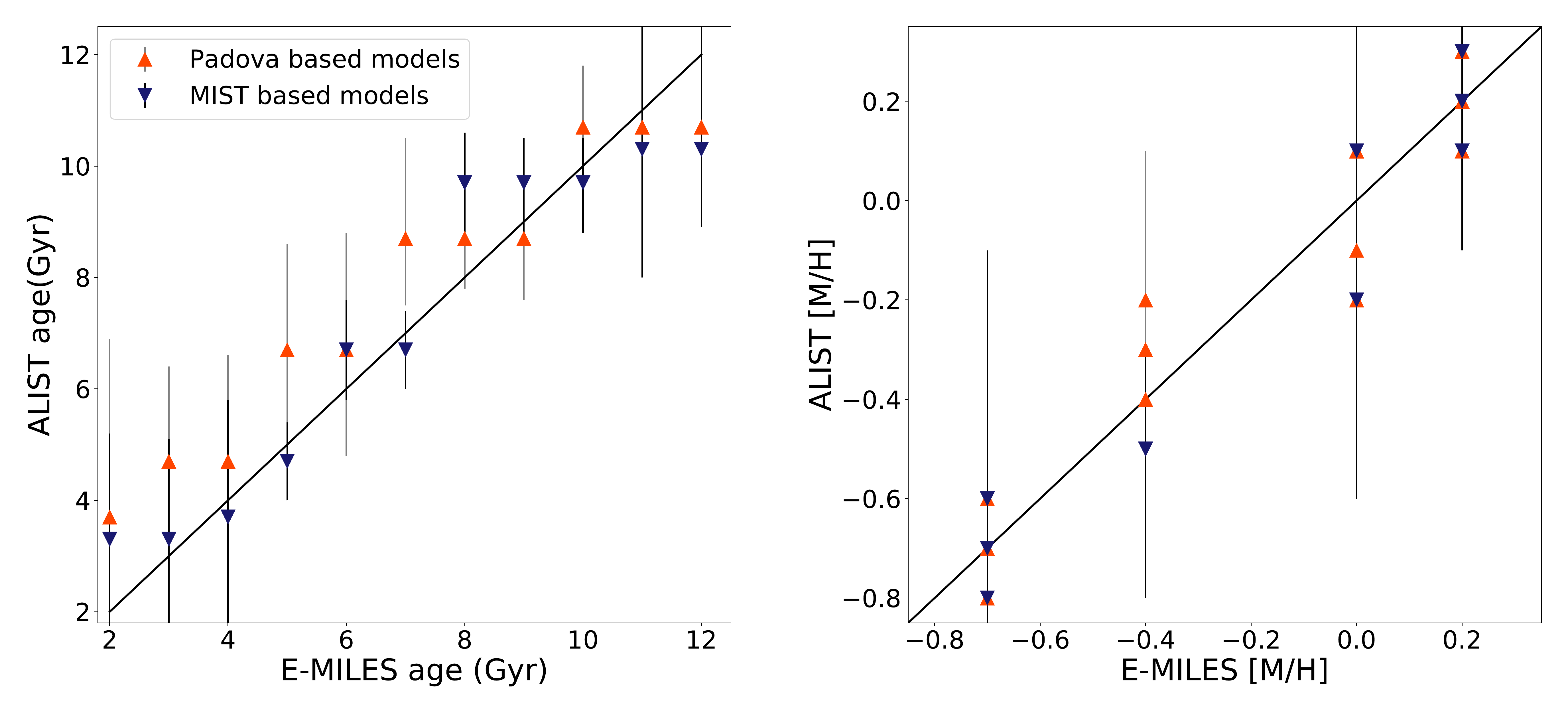}
\caption{Age and [M/H] of best-fit low resolution A-LIST models based on Padova (orange points) and MIST (blue points) isochrones compared with the E-MILES age and [Fe/H]. The uncertainties calculated are explained in Section~\ref{gc}.}
\label{milesprop}
\end{figure*}

\subsection{E-MILES comparison} 
\label{miles}
We also compare our library to the popular E-MILES \citep{vazdekis2016} models for the same age and metallicity. The lower metallicity E-MILES models (i.e, $\rm [Fe/H] < -0.4$) are composite spectra generated by extrapolating the available SSP spectra from MIUSCAT \citep{Vazdekis2012} and the IRTF library \citep[Section 2 of][]{Vazdekis2012}. Since our spectral resolution is much higher than that of  E-MILES, we convolve the A-LIST spectra with a Gaussian kernel to reduce our resolution to match that of the E-MILES spectra. Using the pPXF code and the same fitting method as Section~\ref{gc}, we calculate the likelihood-weighted average values for the age and metallicity. The uncertainty or the ``pseudo-error'' on the A-LIST best fit model is calculated using the same method as explained for the metallicity and $\alpha$--abundance in Section~\ref{gc}. Since these models do not have any errors on them, the pseudo-error is scaled by a normalized error (scatter between best fit model and the data) on the best fit model. 

Figure~\ref{milesprop} shows the computed age and [M/H] from the best-fit reduced resolution Padova-based (orange) and MIST-based (blue) A-LIST templates (with $\rm [\alpha/M]=0$) fitted to E-MILES models at a range of ages (2--12~Gyr) and metallicities (-0.7, -0.4, -0.2, 0.0, and +0.2~dex).

Our models show close correspondence in both age and metallicity to lower resolution E-MILES models. For Padova-based (MIST-based values in parentheses) we find a median offset (E-MILES -- A-LIST) of 0.7 (0.5) Gyr in age and  0.06 (0.06) dex in metallicity, while the median absolute deviation is 1.48 (0.88) Gyr in age and 0.09 (0.17) in metallicity. For the age comparison, MIST-based fits better match the E-MILES models, with both smaller offsets and a smaller median absolute deviation than the Padova-based fits. For metallicities the two models provide quite similar offsets.

\begin{figure}
\includegraphics[trim=2cm 0cm 0cm 0cm, clip, width=0.5\textwidth]{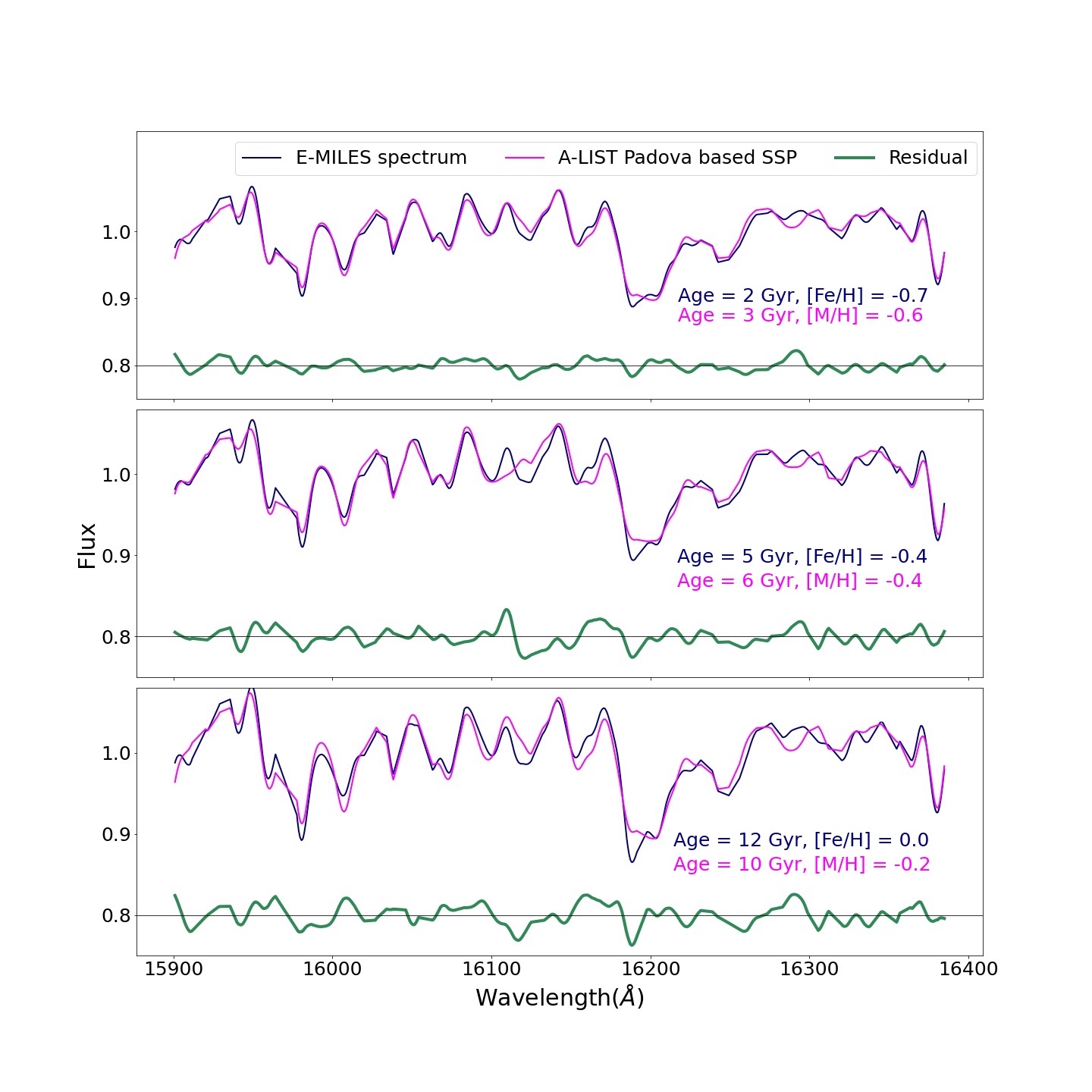}
\caption{The best fit Padova-based A-LIST templates (at reduced resolution, shown in fuchsia) compared with E-MILES spectra (shown in blue) at a range of ages and metallicities (with solar [$\alpha$/M]). The E-MILES age and metallicities are given in each panel in blue, and the best-fitting A-LIST template values in fuchsia. The differences between the spectra (A-LIST -- E-MILES) are shown in green}. The differences in the best-fit values between the Padova-based and MIST-based for these models are small. The fits (Section~\ref{miles}) are performed over the full APOGEE wavelength range, but only a small section is shown here to highlight the similarity.
\label{milescomp}
\end{figure}

Figure~\ref{milescomp} shows comparisons between the best-fit Padova-based A-LIST templates (at the E-MILES resolution; fuchsia) and E-MILES models (blue) using templates with solar-$\alpha$ and a range of ages and metallicities. The differences between the best-fit Padova-based A-LIST and E-MILES models are shown in green.

\subsection{Model spectra quality test based on globular cluster fitting} 
\label{chi2test}

Two major factors affecting the reliability of the A-LIST model spectra, in terms of how representative a spectrum is of its nominal SSP, are the fraction of recovered luminosity (Section~\ref{lumfrac}) and the difference in the luminosity-weighted average T$_\mathrm{eff}$ between the synthetic population and the model spectrum ($\Delta T_\mathrm{eff}$; Section~\ref{avgtemp}). To understand how these affect our spectral models, we conduct tests in which we generate sets of modified A-LIST templates from 
a high luminosity fraction, low $\Delta T_{\rm eff}$ model to simulate lower luminosity fractions and higher $\Delta T_{\rm eff}$ values (e.g., Figure~\ref{hrdlumfrac}). To test the effects of lower fractional luminosities, we remove the bins  that have the largest contributions to the total SSP luminosity. To test the effects of having a non-zero $\Delta T_\mathrm{eff}$, we remove the  hottest and coldest T$_\mathrm{eff}$ bins.

For the recovered luminosity fraction test, we generate these degraded template sets for the best-fitting SSP to three of our high-SNR GCs (B127-G185, B151-G205 and B103-G165, with SNRs of 150, 127 and 103, respectively). We perform the same fitting as in Section~\ref{ppxf} and observe the relation between the $\chi^2$ of the fits and the fractional luminosity. As the fractional luminosity decreases (i.e., the more bins we remove from the HRD), the $\chi^2$ values of the fits increase, as expected. Based on this increase in $\chi^2$, we define a cut-off fractional luminosity for the A-LIST models of 0.32, below which we caution the models may not provide reliable results. This value of 0.32 is selected based on where we see an increase in the reduced $\chi^2$ of $\sim$0.25 relative to the best fit, highest luminosity fraction model.

We note that while this fraction may seem fairly low, many of the spectral bins in the HRD used in constructing our A-LIST models have spectra similar to each other, thus enabling accurate spectral models even at lower fractional luminosities.

We use two approaches for the $\Delta T_\mathrm{eff}$ test. In one, we generate degraded templates to fit the same three high-SNR GCs as above. To test how our models fare at different ages and metallicities, we also create mock data at younger ages and different metallicities by generating noisy data from our models assuming a S/N of 50.  We then fit both the GCs and the mock data to the degraded models using the same fitting method in Section~\ref{ppxf} and observe the relation between the $\chi^2$ of the fits and $\Delta T_\mathrm{eff}$. 
By progressively removing the hottest and coldest T$_\mathrm{eff}$ bins and then fitting the degraded templates to GC spectra or mock data created from our original models, we find that a $\Delta T_\mathrm{eff}$ of $\sim -200$~K and a $\Delta T_\mathrm{eff}$ of $\sim +350$~K correspond to a $\Delta\chi^2 \sim 0.25$.

To summarize, we recommend using only models with luminosity fraction $>$0.32 and $-200~{\rm K} \leq \Delta T_\mathrm{eff} \leq 350$~K; models meeting these requirements have been highlighted in Figure \ref{validplot}. Our ``best'' models (by these metrics) lie in the region of chemical space where the majority of MW stars reside (the right panel of Figure~\ref{apogeeparam}). The quality of our models decreases in bins with few stars available (e.g., regions outside the white line in Figure~\ref{validplot}(a)).
\\
\section{Summary} \label{summary}
In this paper, we present A-LIST: a new spectral library of high resolution, near-infrared integrated-light simple stellar population spectral templates that can be used to compute the properties of complex stellar populations having multiple populations with different ages, [M/H], and [$\alpha$/M]. This library is generated using the isochrone synthesis method with Padova and MIST isochrones and a \citet{Kroupa2001} IMF. We use an empirical stellar library with $\sim$300,000 stars from the APOGEE survey, which has a spectral resolution of $\sim$22,500 and a wavelength range of 15100 to 17000~\AA. The empirical stellar library provides enough APOGEE spectra to generate SSP spectral models representing as much as 99\% of the luminosity of the SSPs, with lower fractional luminosities found in regions of [M/H] and [$\alpha$/M] space that are not well represented in the Milky Way.

Our validation tests demonstrate that our models give fitting results consistent in both age and metallicity with previous measurements and with lower resolution models . First, we fit the A-LIST models to APOGEE spectra of M31 GCs and find metallicities consistent with previous measurements by \citet{Caldwell2011} and \citet{Sakari2016}, with a median offset of $<$0.1~dex.  The [$\alpha$/M] measurements of our M31 GCs are also well-correlated with previously estimated values.  Second, we fit our A-LIST library to E-MILES models at a range of ages and metallicities and find a good agreement with the E-MILES models in the best-fit ages (with a median absolute deviation $<$1.5~Gyr) as well as the best-fit metallicities (with a median absolute deviation $<$0.17 dex).

We also use the fits to the globular clusters and to generated mock data to better understand where our models are most reliable. We find that models with fractional luminosity $\gtrsim$0.32, and with the difference in temperature between the synthetic population and our corresponding model of $-200~{\rm K} \leq \Delta T_\mathrm{eff} \leq 350$~K, provide high quality fits to the spectra. We recommend these thresholds in using our model spectra.

Our models are publicly available at\\
\url{https://github.com/aishashok/ALIST-library}. We expect they will be useful in analyzing medium- and high-resolution $H$-band integrated light spectroscopy of galaxies and star clusters.
\\
\\
\\
\\
\\
\\

We would like to thank Ricardo Schiavon and Nelson Caldwell for providing us with the [Mg/Fe] abundances of M31 GCs. We would also like to thank the anonymous referee whose comments significantly improved this paper. Work on this project by AA, GZ, and AS was supported by NSF grant AST-1911129. Additional work was supported by NSF  PHY-REU Award \#1659494. DAGH acknowledges support from the State Research Agency (AEI) of the Spanish Ministry of Science, Innovation and Universities (MCIU) and the European Regional Development Fund (FEDER) under grant AYA2017-88254-P.

Funding for the Sloan Digital Sky Survey IV has been provided by the Alfred P. Sloan Foundation, the U.S. Department of Energy Office of Science, and the Participating Institutions. SDSS-IV acknowledges support and resources from the Center for High-Performance Computing at the University of Utah. The SDSS web site is www.sdss.org.

SDSS-IV is managed by the Astrophysical Research Consortium for the Participating Institutions of the SDSS Collaboration including the Brazilian Participation Group, the Carnegie Institution for Science, Carnegie Mellon University, the Chilean Participation Group, the French Participation Group, Harvard-Smithsonian Center for Astrophysics, Instituto de Astrof\'isica de Canarias, The Johns Hopkins University, Kavli Institute for the Physics and Mathematics of the Universe (IPMU) / University of Tokyo, the Korean Participation Group, Lawrence Berkeley National Laboratory, Leibniz Institut f\"ur Astrophysik Potsdam (AIP),  Max-Planck-Institut f\"ur Astronomie (MPIA Heidelberg), Max-Planck-Institut f\"ur Astrophysik (MPA Garching), Max-Planck-Institut f\"ur Extraterrestrische Physik (MPE), National Astronomical Observatories of China, New Mexico State University, New York University, University of Notre Dame, Observat\'ario Nacional / MCTI, The Ohio State University, Pennsylvania State University, Shanghai Astronomical Observatory, United Kingdom Participation Group,Universidad Nacional Aut\'onoma de M\'exico, University of Arizona, University of Colorado Boulder, University of Oxford, 
University of Portsmouth, University of Utah, University of Virginia, University of Washington, University of Wisconsin, 
Vanderbilt University, and Yale University.

\bibliography{ALIST}

\begin{thebibliography}{}
\expandafter\ifx\csname natexlab\endcsname\relax\def\natexlab#1{#1}\fi
\providecommand{\url}[1]{\href{#1}{#1}}
\providecommand{\dodoi}[1]{doi:~\href{http://doi.org/#1}{\nolinkurl{#1}}}
\providecommand{\doeprint}[1]{\href{http://ascl.net/#1}{\nolinkurl{http://ascl.net/#1}}}
\providecommand{\doarXiv}[1]{\href{https://arxiv.org/abs/#1}{\nolinkurl{https://arxiv.org/abs/#1}}}

\bibitem[{{Ahumada} {et~al.}(2019){Ahumada}, {Allende Prieto}, {Almeida},
  {Anders}, {Anderson}, {Andrews}, {Anguiano}, {Arcodia}, {Armengaud},
  {Aubert}, {Avila}, {Avila-Reese}, {Badenes}, {Balland}, {Barger},
  {Barrera-Ballesteros}, {Basu}, {Bautista}, {Beaton}, {Beers}, {Benavides},
  {Bender}, {Bernardi}, {Bershady}, {Beutler}, {Moni Bidin}, {Bird}, {Bizyaev},
  {Blanc}, {Blanton}, {Boquien}, {Borissova}, {Bovy}, {Brandt}, {Brinkmann},
  {Brownstein}, {Bundy}, {Bureau}, {Burgasser}, {Burtin}, {Cano-Diaz},
  {Capasso}, {Cappellari}, {Carrera}, {Chabanier}, {Chaplin}, {Chapman},
  {Cherinka}, {Chiappini}, {Choi}, {Chojnowski}, {Chung}, {Clerc}, {Coffey},
  {Comerford}, {Comparat}, {da Costa}, {Cousinou}, {Covey}, {Crane}, {Cunha},
  {da Silva Ilha}, {Dai}, {Damsted}, {Darling}, {Horta Darrington}, {Davidson},
  {Davies}, {Dawson}, {De}, {de la Macorra}, {De Lee}, {Queiroz}, {Deconto
  Machado}, {de la Torre}, {Dell'Agli}, {du Mas des Bourboux},
  {Diamond-Stanic}, {Dillon}, {Donor}, {Drory}, {Duckworth}, {Dwelly},
  {Ebelke}, {Eftekharzadeh}, {Davis Eigenbrot}, {Elsworth}, {Eracleous},
  {Erfanianfar}, {Escoffier}, {Fan}, {Farr}, {Fernandez-Trincado}, {Feuillet},
  {Finoguenov}, {Fofie}, {Fraser-McKelvie}, {Frinchaboy}, {Fromenteau}, {Fu},
  {Galbany}, {Garcia}, {Garcia-Hernandez}, {Garma Oehmichen}, {Ge}, {Geimba
  Maia}, {Geisler}, {Gelfand }, {Goddy}, {Le Goff}, {Gonzalez-Perez},
  {Grabowski}, {Green}, {Grier}, {Guo}, {Guy}, {Harding}, {Hasselquist},
  {Hawken}, {Hayes}, {Hearty}, {Hekker}, {Hogg}, {Holtzman}, {Hou}, {Hsieh},
  {Huber}, {Hunt}, {Ider Chitham}, {Imig}, {Jaber}, {Jimenez Angel}, {Johnson},
  {Jones}, {Jonsson}, {Jullo}, {Kim}, {Kinemuchi}, {Kirkpatrick}, {Kite},
  {Klaene}, {Kneib}, {Kollmeier}, {Kong}, {Kounkel}, {Krishnarao}, {Lacerna},
  {Lan}, {Lane}, {Law}, {Leung}, {Lewis}, {Li}, {Lian}, {Lin}, {Long},
  {Longa-Pena}, {Lundgren}, {Lyke}, {Mackereth}, {MacLeod}, {Majewski},
  {Manchado}, {Maraston}, {Martini}, {Masseron}, {Masters}, {Mathur},
  {McDermid}, {Merloni}, {Merrifield}, {Meszaros}, {Miglio}, {Minniti},
  {Minsley}, {Miyaji}, {Gohar Mohammad}, {Mosser}, {Mueller}, {Muna},
  {Munoz-Gutierrez}, {Myers}, {Nadathur}, {Nair}, {Correa do Nascimento},
  {Nevin}, {Newman}, {Nidever}, {Nitschelm}, {Noterdaeme}, {O'Connell},
  {Olmstead}, {Oravetz}, {Oravetz}, {Osorio}, {Pace}, {Padilla},
  {Palanque-Delabrouille}, {Palicio}, {Pan}, {Pan}, {Parker}, {Paviot},
  {Peirani}, {Pena Ramrez}, {Penny}, {Percival}, {Perez-Fournon},
  {Perez-Rafols}, {Petitjean}, {Pieri}, {Pinsonneault}, {Poovelil}, {Povick},
  {Prakash}, {Price-Whelan}, {Raddick}, {Raichoor}, {Ray}, {Barboza Rembold},
  {Rezaie}, {Riffel}, {Riffel}, {Rix}, {Robin}, {Roman-Lopes}, {Roman-Zuniga},
  {Rose}, {Ross}, {Rossi}, {Rowlands}, {Rubin}, {Salvato}, {Sanchez},
  {Sanchez-Menguiano}, {Sanchez-Gallego}, {Sayres}, {Schaefer}, {Schiavon},
  {Schimoia}, {Schlafly}, {Schlegel}, {Schneider}, {Schultheis}, {Schwope},
  {Seo}, {Serenelli}, {Shafieloo}, {Shamsi}, {Shao}, {Shen}, {Shetrone},
  {Shirley}, {Silva Aguirre}, {Simon}, {Skrutskie}, {Slosar}, {Smethurst},
  {Sobeck}, {Cervantes Sodi}, {Souto}, {Stark}, {Stassun}, {Steinmetz},
  {Stello}, {Stermer}, {Storchi-Bergmann}, {Streblyanska}, {Stringfellow},
  {Stutz}, {Suarez}, {Sun}, {Taghizadeh-Popp}, {Talbot}, {Tayar}, {Thakar},
  {Theriault}, {Thomas}, {Thomas}, {Tinker}, {Tojeiro}, {Hernandez Toledo},
  {Tremonti}, {Troup}, {Tuttle}, {Unda-Sanzana}, {Valentini},
  {Vargas-Gonzalez}, {Vargas-Magana}, {Vazquez-Mata}, {Vivek}, {Wake}, {Wang},
  {Weaver}, {Weijmans}, {Wild}, {Wilson}, {Wilson}, {Wolthuis}, {Wood-Vasey},
  {Yan}, {Yang}, {Yeche}, {Zamora}, {Zarrouk}, {Zasowski}, {Zhang}, {Zhao},
  {Zhao}, {Zheng}, {Zheng}, {Zhu}, \& {Zou}}]{Ahumada2019}
{Ahumada}, R., {Allende Prieto}, C., {Almeida}, A., {et~al.} 2019, arXiv
  e-prints, arXiv:1912.02905.
\newblock \doarXiv{1912.02905}

\bibitem[{{Baldwin} {et~al.}(2018){Baldwin}, {McDermid}, {Kuntschner},
  {Maraston}, \& {Conroy}}]{Baldwin2018}
{Baldwin}, C., {McDermid}, R.~M., {Kuntschner}, H., {Maraston}, C., \&
  {Conroy}, C. 2018, \mnras, 473, 4698, \dodoi{10.1093/mnras/stx2502}

\bibitem[{{Beerman} {et~al.}(2012){Beerman}, {Johnson}, {Fouesneau},
  {Dalcanton}, {Weisz}, {Seth}, {Williams}, {Bell}, {Bianchi}, {Caldwell},
  {Dolphin}, {Gouliermis}, {Kalirai}, {Larsen}, {Melbourne}, {Rix}, \&
  {Skillman}}]{Beerman2012}
{Beerman}, L.~C., {Johnson}, L.~C., {Fouesneau}, M., {et~al.} 2012, \apj, 760,
  104, \dodoi{10.1088/0004-637X/760/2/104}

\bibitem[{{Blanton} {et~al.}(2017){Blanton}, {Bershady}, {Abolfathi},
  {Albareti}, {Allende Prieto}, {Almeida}, {Alonso-Garc{\'\i}a}, {Anders},
  {Anderson}, {Andrews}, {Aquino-Ort{\'\i}z}, {Arag{\'o}n-Salamanca},
  {Argudo-Fern{\'a}ndez}, {Armengaud}, {Aubourg}, {Avila-Reese}, {Badenes},
  {Bailey}, {Barger}, {Barrera-Ballesteros}, {Bartosz}, {Bates}, {Baumgarten},
  {Bautista}, {Beaton}, {Beers}, {Belfiore}, {Bender}, {Berlind}, {Bernardi},
  {Beutler}, {Bird}, {Bizyaev}, {Blanc}, {Blomqvist}, {Bolton}, {Boquien},
  {Borissova}, {van den Bosch}, {Bovy}, {Brandt}, {Brinkmann}, {Brownstein},
  {Bundy}, {Burgasser}, {Burtin}, {Busca}, {Cappellari}, {Delgado Carigi},
  {Carlberg}, {Carnero Rosell}, {Carrera}, {Chanover}, {Cherinka}, {Cheung},
  {G{\'o}mez Maqueo Chew}, {Chiappini}, {Choi}, {Chojnowski}, {Chuang},
  {Chung}, {Cirolini}, {Clerc}, {Cohen}, {Comparat}, {da Costa}, {Cousinou},
  {Covey}, {Crane}, {Croft}, {Cruz-Gonzalez}, {Garrido Cuadra}, {Cunha},
  {Damke}, {Darling}, {Davies}, {Dawson}, {de la Macorra}, {Dell'Agli}, {De
  Lee}, {Delubac}, {Di Mille}, {Diamond-Stanic}, {Cano-D{\'\i}az}, {Donor},
  {Downes}, {Drory}, {du Mas des Bourboux}, {Duckworth}, {Dwelly}, {Dyer},
  {Ebelke}, {Eigenbrot}, {Eisenstein}, {Emsellem}, {Eracleous}, {Escoffier},
  {Evans}, {Fan}, {Fern{\'a}ndez-Alvar}, {Fernandez-Trincado}, {Feuillet},
  {Finoguenov}, {Fleming}, {Font-Ribera}, {Fredrickson}, {Freischlad},
  {Frinchaboy}, {Fuentes}, {Galbany}, {Garcia-Dias},
  {Garc{\'\i}a-Hern{\'a}ndez}, {Gaulme}, {Geisler}, {Gelfand},
  {Gil-Mar{\'\i}n}, {Gillespie}, {Goddard}, {Gonzalez-Perez}, {Grabowski},
  {Green}, {Grier}, {Gunn}, {Guo}, {Guy}, {Hagen}, {Hahn}, {Hall}, {Harding},
  {Hasselquist}, {Hawley}, {Hearty}, {Gonzalez Hern{\'a}ndez}, {Ho}, {Hogg},
  {Holley-Bockelmann}, {Holtzman}, {Holzer}, {Huehnerhoff}, {Hutchinson},
  {Hwang}, {Ibarra-Medel}, {da Silva Ilha}, {Ivans}, {Ivory}, {Jackson},
  {Jensen}, {Johnson}, {Jones}, {J{\"o}nsson}, {Jullo}, {Kamble}, {Kinemuchi},
  {Kirkby}, {Kitaura}, {Klaene}, {Knapp}, {Kneib}, {Kollmeier}, {Lacerna},
  {Lane}, {Lang}, {Law}, {Lazarz}, {Lee}, {Le Goff}, {Liang}, {Li}, {Li},
  {Lian}, {Lima}, {Lin}, {Lin}, {Bertran de Lis}, {Liu}, {de Icaza Lizaola},
  {Long}, {Lucatello}, {Lundgren}, {MacDonald}, {Deconto Machado}, {MacLeod},
  {Mahadevan}, {Geimba Maia}, {Maiolino}, {Majewski}, {Malanushenko},
  {Malanushenko}, {Manchado}, {Mao}, {Maraston}, {Marques-Chaves}, {Masseron},
  {Masters}, {McBride}, {McDermid}, {McGrath}, {McGreer}, {Medina Pe{\~n}a},
  {Melendez}, {Merloni}, {Merrifield}, {Meszaros}, {Meza}, {Minchev},
  {Minniti}, {Miyaji}, {More}, {Mulchaey}, {M{\"u}ller-S{\'a}nchez}, {Muna},
  {Munoz}, {Myers}, {Nair}, {Nandra}, {Correa do Nascimento}, {Negrete},
  {Ness}, {Newman}, {Nichol}, {Nidever}, {Nitschelm}, {Ntelis}, {O'Connell},
  {Oelkers}, {Oravetz}, {Oravetz}, {Pace}, {Padilla}, {Palanque-Delabrouille},
  {Alonso Palicio}, {Pan}, {Parejko}, {Parikh}, {P{\^a}ris}, {Park}, {Patten},
  {Peirani}, {Pellejero-Ibanez}, {Penny}, {Percival}, {Perez-Fournon},
  {Petitjean}, {Pieri}, {Pinsonneault}, {Pisani}, {Poleski}, {Prada},
  {Prakash}, {Queiroz}, {Raddick}, {Raichoor}, {Barboza Rembold}, {Richstein},
  {Riffel}, {Riffel}, {Rix}, {Robin}, {Rockosi}, {Rodr{\'\i}guez-Torres},
  {Roman-Lopes}, {Rom{\'a}n-Z{\'u}{\~n}iga}, {Rosado}, {Ross}, {Rossi}, {Ruan},
  {Ruggeri}, {Rykoff}, {Salazar-Albornoz}, {Salvato}, {S{\'a}nchez}, {Aguado},
  {S{\'a}nchez-Gallego}, {Santana}, {Santiago}, {Sayres}, {Schiavon}, {da Silva
  Schimoia}, {Schlafly}, {Schlegel}, {Schneider}, {Schultheis}, {Schuster},
  {Schwope}, {Seo}, {Shao}, {Shen}, {Shetrone}, {Shull}, {Simon}, {Skinner},
  {Skrutskie}, {Slosar}, {Smith}, {Sobeck}, {Sobreira}, {Somers}, {Souto},
  {Stark}, {Stassun}, {Stauffer}, {Steinmetz}, {Storchi-Bergmann},
  {Streblyanska}, {Stringfellow}, {Su{\'a}rez}, {Sun}, {Suzuki}, {Szigeti},
  {Taghizadeh-Popp}, {Tang}, {Tao}, {Tayar}, {Tembe}, {Teske}, {Thakar},
  {Thomas}, {Thompson}, {Tinker}, {Tissera}, {Tojeiro}, {Hernandez Toledo}, {de
  la Torre}, {Tremonti}, {Troup}, {Valenzuela}, {Martinez Valpuesta},
  {Vargas-Gonz{\'a}lez}, {Vargas-Maga{\~n}a}, {Vazquez}, {Villanova}, {Vivek},
  {Vogt}, {Wake}, {Walterbos}, {Wang}, {Weaver}, {Weijmans}, {Weinberg},
  {Westfall}, {Whelan}, {Wild}, {Wilson}, {Wood-Vasey}, {Wylezalek}, {Xiao},
  {Yan}, {Yang}, {Ybarra}, {Y{\`e}che}, {Zakamska}, {Zamora}, {Zarrouk},
  {Zasowski}, {Zhang}, {Zhao}, {Zheng}, {Zheng}, {Zhou}, {Zhou}, {Zhu},
  {Zoccali}, \& {Zou}}]{Blanton2017}
{Blanton}, M.~R., {Bershady}, M.~A., {Abolfathi}, B., {et~al.} 2017, \aj, 154,
  28, \dodoi{10.3847/1538-3881/aa7567}

\bibitem[{{Boardman} {et~al.}(2020){Boardman}, {Zasowski}, {Seth}, {Newman},
  {Andrews}, {Bershady}, {Bird}, {Chiappini}, {Fielder}, {Fraser-McKelvie},
  {Jones}, {Licquia}, {Masters}, {Minchev}, {Schiavon}, {Brownstein}, {Drory},
  \& {Lane}}]{Boardman2020}
{Boardman}, N., {Zasowski}, G., {Seth}, A., {et~al.} 2020, \mnras, 491, 3672,
  \dodoi{10.1093/mnras/stz3126}

\bibitem[{{B{\"o}ker} {et~al.}(1999){B{\"o}ker}, {van der Marel}, \&
  {Vacca}}]{Boker1999}
{B{\"o}ker}, T., {van der Marel}, R.~P., \& {Vacca}, W.~D. 1999, \aj, 118, 831,
  \dodoi{10.1086/300985}

\bibitem[{Bowen \& Vaughan(1973)}]{Bowen1973}
Bowen, I.~S., \& Vaughan, A.~H. 1973, Appl. Opt., 12, 1430,
  \dodoi{10.1364/AO.12.001430}

\bibitem[{{Bressan} {et~al.}(2012){Bressan}, {Marigo}, {Girardi}, {Salasnich},
  {Dal Cero}, {Rubele}, \& {Nanni}}]{Bressan2012}
{Bressan}, A., {Marigo}, P., {Girardi}, L., {et~al.} 2012, \mnras, 427, 127,
  \dodoi{10.1111/j.1365-2966.2012.21948.x}

\bibitem[{Bruzual \& Charlot(2003)}]{Bruzal2003}
Bruzual, G., \& Charlot, S. 2003, Monthly Notices of the Royal Astronomical
  Society, 344, 1000, \dodoi{10.1046/j.1365-8711.2003.06897.x}

\bibitem[{{Caldwell} {et~al.}(2011){Caldwell}, {Schiavon}, {Morrison}, {Rose},
  \& {Harding}}]{Caldwell2011}
{Caldwell}, N., {Schiavon}, R., {Morrison}, H., {Rose}, J.~A., \& {Harding}, P.
  2011, \aj, 141, 61, \dodoi{10.1088/0004-6256/141/2/61}

\bibitem[{{Cappellari}(2017)}]{Cappellari2017}
{Cappellari}, M. 2017, \mnras, 466, 798, \dodoi{10.1093/mnras/stw3020}

\bibitem[{{Cappellari} {et~al.}(2009){Cappellari}, {di Serego Alighieri},
  {Cimatti}, {Daddi}, {Renzini}, {Kurk}, {Cassata}, {Dickinson},
  {Franceschini}, {Mignoli}, {Pozzetti}, {Rodighiero}, {Rosati}, \&
  {Zamorani}}]{Cappellari2009}
{Cappellari}, M., {di Serego Alighieri}, S., {Cimatti}, A., {et~al.} 2009,
  \apjl, 704, L34, \dodoi{10.1088/0004-637X/704/1/L34}

\bibitem[{{Cenarro} {et~al.}(2001){Cenarro}, {Cardiel}, {Gorgas}, {Peletier},
  {Vazdekis}, \& {Prada}}]{Cenarro2001}
{Cenarro}, A.~J., {Cardiel}, N., {Gorgas}, J., {et~al.} 2001, \mnras, 326, 959,
  \dodoi{10.1046/j.1365-8711.2001.04688.x}

\bibitem[{{Cenarro} {et~al.}(2009){Cenarro}, {Cardiel}, {Vazdekis}, \&
  {Gorgas}}]{Cenarro2009}
{Cenarro}, A.~J., {Cardiel}, N., {Vazdekis}, A., \& {Gorgas}, J. 2009, \mnras,
  396, 1895, \dodoi{10.1111/j.1365-2966.2009.14839.x}

\bibitem[{{Cesetti} {et~al.}(2009){Cesetti}, {Ivanov}, {Morelli}, {Pizzella},
  {Buson}, {Corsini}, {Dalla Bont{\`a}}, {Stiavelli}, \&
  {Bertola}}]{cesetti2009}
{Cesetti}, M., {Ivanov}, V.~D., {Morelli}, L., {et~al.} 2009, \aap, 497, 41,
  \dodoi{10.1051/0004-6361/200810506}

\bibitem[{{Chavez} {et~al.}(1996){Chavez}, {Malagnini}, \&
  {Morossi}}]{Chavez1996}
{Chavez}, M., {Malagnini}, M.~L., \& {Morossi}, C. 1996, \apj, 471, 726,
  \dodoi{10.1086/178001}

\bibitem[{{Chen} {et~al.}(2015){Chen}, {Bressan}, {Girardi}, {Marigo}, {Kong},
  \& {Lanza}}]{Chen2015}
{Chen}, Y., {Bressan}, A., {Girardi}, L., {et~al.} 2015, \mnras, 452, 1068,
  \dodoi{10.1093/mnras/stv1281}

\bibitem[{{Chen} {et~al.}(2014){Chen}, {Girardi}, {Bressan}, {Marigo},
  {Barbieri}, \& {Kong}}]{Chen2014}
{Chen}, Y., {Girardi}, L., {Bressan}, A., {et~al.} 2014, \mnras, 444, 2525,
  \dodoi{10.1093/mnras/stu1605}

\bibitem[{{Choi} {et~al.}(2016){Choi}, {Dotter}, {Conroy}, {Cantiello},
  {Paxton}, \& {Johnson}}]{Choi2016}
{Choi}, J., {Dotter}, A., {Conroy}, C., {et~al.} 2016, \apj, 823, 102,
  \dodoi{10.3847/0004-637X/823/2/102}

\bibitem[{{Cid Fernandes} {et~al.}(2011){Cid Fernandes}, {Mateus}, {Sodr{\'e}},
  {Stasinska}, \& {Gomes}}]{cidfernandes2011}
{Cid Fernandes}, R., {Mateus}, A., {Sodr{\'e}}, L., {Stasinska}, G., \&
  {Gomes}, J.~M. 2011, {STARLIGHT: Spectral Synthesis Code}.
\newblock \doeprint{1108.006}

\bibitem[{{Cignoni} {et~al.}(2019){Cignoni}, {Sacchi}, {Tosi}, {Aloisi},
  {Cook}, {Calzetti}, {Lee}, {Sabbi}, {Thilker}, {Adamo}, {Dale}, {Elmegreen},
  {Gallagher}, {Grebel}, {Johnson}, {Messa}, {Smith}, \& {Ubeda}}]{Cignoni2019}
{Cignoni}, M., {Sacchi}, E., {Tosi}, M., {et~al.} 2019, \apj, 887, 112,
  \dodoi{10.3847/1538-4357/ab53d5}

\bibitem[{{Cirasuolo} \& {MOONS Consortium}(2016)}]{MOONS}
{Cirasuolo}, M., \& {MOONS Consortium}. 2016, Astronomical Society of the
  Pacific Conference Series, Vol. 507, {MOONS: A New Powerful Multi-Object
  Spectrograph for the VLT}, ed. I.~{Skillen}, M.~{Balcells}, \& S.~{Trager},
  109

\bibitem[{{Coelho} {et~al.}(2020){Coelho}, {Bruzual}, \&
  {Charlot}}]{Coelho2020}
{Coelho}, P. R.~T., {Bruzual}, G., \& {Charlot}, S. 2020, \mnras, 491, 2025,
  \dodoi{10.1093/mnras/stz3023}

\bibitem[{{Conroy}(2013)}]{conroy2013}
{Conroy}, C. 2013, \araa, 51, 393, \dodoi{10.1146/annurev-astro-082812-141017}

\bibitem[{{Conroy} {et~al.}(2018){Conroy}, {Villaume}, {van Dokkum}, \&
  {Lind}}]{Conroy2018}
{Conroy}, C., {Villaume}, A., {van Dokkum}, P.~G., \& {Lind}, K. 2018, \apj,
  854, 139, \dodoi{10.3847/1538-4357/aaab49}

\bibitem[{{Cushing} {et~al.}(2005){Cushing}, {Rayner}, \&
  {Vacca}}]{Cushing2005}
{Cushing}, M.~C., {Rayner}, J.~T., \& {Vacca}, W.~D. 2005, \apj, 623, 1115,
  \dodoi{10.1086/428040}

\bibitem[{{Dahmer-Hahn} {et~al.}(2018){Dahmer-Hahn}, {Riffel},
  {Rodr{\'\i}guez-Ardila}, {Martins}, {Kehrig}, {Heckman}, {Pastoriza}, \&
  {Dametto}}]{Dahmer-Hahn2018}
{Dahmer-Hahn}, L.~G., {Riffel}, R., {Rodr{\'\i}guez-Ardila}, A., {et~al.} 2018,
  \mnras, 476, 4459, \dodoi{10.1093/mnras/sty515}

\bibitem[{{De Silva} {et~al.}(2015){De Silva}, {Freeman}, {Bland-Hawthorn},
  {Martell}, {de Boer}, {Asplund}, {Keller}, {Sharma}, {Zucker}, {Zwitter},
  {Anguiano}, {Bacigalupo}, {Bayliss}, {Beavis}, {Bergemann}, {Campbell},
  {Cannon}, {Carollo}, {Casagrande}, {Casey}, {Da Costa}, {D'Orazi}, {Dotter},
  {Duong}, {Heger}, {Ireland}, {Kafle}, {Kos}, {Lattanzio}, {Lewis}, {Lin},
  {Lind}, {Munari}, {Nataf}, {O'Toole}, {Parker}, {Reid}, {Schlesinger},
  {Sheinis}, {Simpson}, {Stello}, {Ting}, {Traven}, {Watson}, {Wittenmyer},
  {Yong}, \& {{\v{Z}}erjal}}]{DeSilva2015}
{De Silva}, G.~M., {Freeman}, K.~C., {Bland-Hawthorn}, J., {et~al.} 2015,
  \mnras, 449, 2604, \dodoi{10.1093/mnras/stv327}

\bibitem[{{Dotter}(2016)}]{Dotter2016}
{Dotter}, A. 2016, \apjs, 222, 8, \dodoi{10.3847/0067-0049/222/1/8}

\bibitem[{{Durbin} {et~al.}(2020){Durbin}, {Beaton}, {Dalcanton}, {Williams},
  \& {Boyer}}]{Durbin2020}
{Durbin}, M.~J., {Beaton}, R.~L., {Dalcanton}, J.~J., {Williams}, B.~F., \&
  {Boyer}, M.~L. 2020, arXiv e-prints, arXiv:2006.08559.
\newblock \doarXiv{2006.08559}

\bibitem[{{Eisenhauer} {et~al.}(2003){Eisenhauer}, {Abuter}, {Bickert},
  {Biancat-Marchet}, {Bonnet}, {Brynnel}, {Conzelmann}, {Delabre}, {Donaldson},
  {Farinato}, {Fedrigo}, {Genzel}, {Hubin}, {Iserlohe}, {Kasper},
  {Kissler-Patig}, {Monnet}, {Roehrle}, {Schreiber}, {Stroebele}, {Tecza},
  {Thatte}, \& {Weisz}}]{SINFONI}
{Eisenhauer}, F., {Abuter}, R., {Bickert}, K., {et~al.} 2003, Society of
  Photo-Optical Instrumentation Engineers (SPIE) Conference Series, Vol. 4841,
  {SINFONI - Integral field spectroscopy at 50 milli-arcsecond resolution with
  the ESO VLT}, ed. M.~{Iye} \& A.~F.~M. {Moorwood}, 1548--1561

\bibitem[{{Eisenstein} {et~al.}(2011){Eisenstein}, {Weinberg}, {Agol},
  {Aihara}, {Allende Prieto}, {Anderson}, {Arns}, {Aubourg}, {Bailey},
  {Balbinot}, {Barkhouser}, {Beers}, {Berlind}, {Bickerton}, {Bizyaev},
  {Blanton}, {Bochanski}, {Bolton}, {Bosman}, {Bovy}, {Brandt}, {Breslauer},
  {Brewington}, {Brinkmann}, {Brown}, {Brownstein}, {Burger}, {Busca},
  {Campbell}, {Cargile}, {Carithers}, {Carlberg}, {Carr}, {Chang}, {Chen},
  {Chiappini}, {Comparat}, {Connolly}, {Cortes}, {Croft}, {Cunha}, {da Costa},
  {Davenport}, {Dawson}, {De Lee}, {Porto de Mello}, {de Simoni}, {Dean},
  {Dhital}, {Ealet}, {Ebelke}, {Edmondson}, {Eiting}, {Escoffier}, {Esposito},
  {Evans}, {Fan}, {Femen{\'\i}a Castell{\'a}}, {Dutra Ferreira}, {Fitzgerald},
  {Fleming}, {Font-Ribera}, {Ford}, {Frinchaboy}, {Garc{\'\i}a P{\'e}rez},
  {Gaudi}, {Ge}, {Ghezzi}, {Gillespie}, {Gilmore}, {Girardi}, {Gott}, {Gould},
  {Grebel}, {Gunn}, {Hamilton}, {Harding}, {Harris}, {Hawley}, {Hearty},
  {Hennawi}, {Gonz{\'a}lez Hern{\'a}ndez}, {Ho}, {Hogg}, {Holtzman},
  {Honscheid}, {Inada}, {Ivans}, {Jiang}, {Jiang}, {Johnson}, {Jordan},
  {Jordan}, {Kauffmann}, {Kazin}, {Kirkby}, {Klaene}, {Knapp}, {Kneib},
  {Kochanek}, {Koesterke}, {Kollmeier}, {Kron}, {Lampeitl}, {Lang}, {Lawler},
  {Le Goff}, {Lee}, {Lee}, {Leisenring}, {Lin}, {Liu}, {Long}, {Loomis},
  {Lucatello}, {Lundgren}, {Lupton}, {Ma}, {Ma}, {MacDonald}, {Mack},
  {Mahadevan}, {Maia}, {Majewski}, {Makler}, {Malanushenko}, {Malanushenko},
  {Mand elbaum}, {Maraston}, {Margala}, {Maseman}, {Masters}, {McBride},
  {McDonald}, {McGreer}, {McMahon}, {Mena Requejo}, {M{\'e}nard},
  {Miralda-Escud{\'e}}, {Morrison}, {Mullally}, {Muna}, {Murayama}, {Myers},
  {Naugle}, {Neto}, {Nguyen}, {Nichol}, {Nidever}, {O'Connell}, {Ogando},
  {Olmstead}, {Oravetz}, {Padmanabhan}, {Paegert}, {Palanque-Delabrouille},
  {Pan}, {Pandey}, {Parejko}, {P{\^a}ris}, {Pellegrini}, {Pepper}, {Percival},
  {Petitjean}, {Pfaffenberger}, {Pforr}, {Phleps}, {Pichon}, {Pieri}, {Prada},
  {Price-Whelan}, {Raddick}, {Ramos}, {Reid}, {Reyle}, {Rich}, {Richards},
  {Rieke}, {Rieke}, {Rix}, {Robin}, {Rocha-Pinto}, {Rockosi}, {Roe},
  {Rollinde}, {Ross}, {Ross}, {Rossetto}, {S{\'a}nchez}, {Santiago}, {Sayres},
  {Schiavon}, {Schlegel}, {Schlesinger}, {Schmidt}, {Schneider}, {Sellgren},
  {Shelden}, {Sheldon}, {Shetrone}, {Shu}, {Silverman}, {Simmerer}, {Simmons},
  {Sivarani}, {Skrutskie}, {Slosar}, {Smee}, {Smith}, {Snedden}, {Stassun},
  {Steele}, {Steinmetz}, {Stockett}, {Stollberg}, {Strauss}, {Szalay},
  {Tanaka}, {Thakar}, {Thomas}, {Tinker}, {Tofflemire}, {Tojeiro}, {Tremonti},
  {Vargas Maga{\~n}a}, {Verde}, {Vogt}, {Wake}, {Wan}, {Wang}, {Weaver},
  {White}, {White}, {Wilson}, {Wisniewski}, {Wood-Vasey}, {Yanny}, {Yasuda},
  {Y{\`e}che}, {York}, {Young}, {Zasowski}, {Zehavi}, \&
  {Zhao}}]{Eisenstein2011}
{Eisenstein}, D.~J., {Weinberg}, D.~H., {Agol}, E., {et~al.} 2011, \aj, 142,
  72, \dodoi{10.1088/0004-6256/142/3/72}

\bibitem[{Elias {et~al.}(2006)Elias, Joyce, Liang, Muller, Hileman, \&
  George}]{GNIRS}
Elias, J.~H., Joyce, R.~R., Liang, M., {et~al.} 2006, in Ground-based and
  Airborne Instrumentation for Astronomy, ed. I.~S. McLean \& M.~Iye, Vol.
  6269, International Society for Optics and Photonics (SPIE), 1389 -- 1400.
\newblock \url{https://doi.org/10.1117/12.671817}

\bibitem[{{F{\"o}rster Schreiber} {et~al.}(2003){F{\"o}rster Schreiber},
  {Genzel}, {Lutz}, \& {Sternberg}}]{Forster2003}
{F{\"o}rster Schreiber}, N.~M., {Genzel}, R., {Lutz}, D., \& {Sternberg}, A.
  2003, \apj, 599, 193, \dodoi{10.1086/379097}

\bibitem[{{Gallazzi} {et~al.}(2005){Gallazzi}, {Charlot}, {Brinchmann},
  {White}, \& {Tremonti}}]{Gallazzi2005}
{Gallazzi}, A., {Charlot}, S., {Brinchmann}, J., {White}, S. D.~M., \&
  {Tremonti}, C.~A. 2005, \mnras, 362, 41,
  \dodoi{10.1111/j.1365-2966.2005.09321.x}

\bibitem[{{Garc{\'\i}a P{\'e}rez} {et~al.}(2016){Garc{\'\i}a P{\'e}rez},
  {Allende Prieto}, {Holtzman}, {Shetrone}, {M{\'e}sz{\'a}ros}, {Bizyaev},
  {Carrera}, {Cunha}, {Garc{\'\i}a-Hern{\'a}ndez}, {Johnson}, {Majewski},
  {Nidever}, {Schiavon}, {Shane}, {Smith}, {Sobeck}, {Troup}, {Zamora},
  {Weinberg}, {Bovy}, {Eisenstein}, {Feuillet}, {Frinchaboy}, {Hayden},
  {Hearty}, {Nguyen}, {O'Connell}, {Pinsonneault}, {Wilson}, \&
  {Zasowski}}]{Garcia2016}
{Garc{\'\i}a P{\'e}rez}, A.~E., {Allende Prieto}, C., {Holtzman}, J.~A.,
  {et~al.} 2016, \aj, 151, 144, \dodoi{10.3847/0004-6256/151/6/144}

\bibitem[{{Gonneau} {et~al.}(2020){Gonneau}, {Lyubenova}, {Lan{\c{c}}on},
  {Trager}, {Peletier}, {Arentsen}, {Chen}, {Coelho}, {Dries},
  {Falc{\'o}n-Barroso}, {Prugniel}, {S{\'a}nchez-Bl{\'a}zquez}, {Vazdekis}, \&
  {Verro}}]{Gonneau2020}
{Gonneau}, A., {Lyubenova}, M., {Lan{\c{c}}on}, A., {et~al.} 2020, \aap, 634,
  A133, \dodoi{10.1051/0004-6361/201936825}

\bibitem[{{Gratton} {et~al.}(2012){Gratton}, {Carretta}, \&
  {Bragaglia}}]{Gratton2012}
{Gratton}, R.~G., {Carretta}, E., \& {Bragaglia}, A. 2012, \aapr, 20, 50,
  \dodoi{10.1007/s00159-012-0050-3}

\bibitem[{{Gunn} {et~al.}(2006){Gunn}, {Siegmund}, {Mannery}, {Owen}, {Hull},
  {Leger}, {Carey}, {Knapp}, {York}, {Boroski}, {Kent}, {Lupton}, {Rockosi},
  {Evans}, {Waddell}, {Anderson}, {Annis}, {Barentine}, {Bartoszek}, {Bastian},
  {Bracker}, {Brewington}, {Briegel}, {Brinkmann}, {Brown}, {Carr},
  {Czarapata}, {Drennan}, {Dombeck}, {Federwitz}, {Gillespie}, {Gonzales},
  {Hansen}, {Harvanek}, {Hayes}, {Jordan}, {Kinney}, {Klaene}, {Kleinman},
  {Kron}, {Kresinski}, {Lee}, {Limmongkol}, {Lindenmeyer}, {Long}, {Loomis},
  {McGehee}, {Mantsch}, {Neilsen}, {Neswold}, {Newman}, {Nitta}, {Peoples},
  {Pier}, {Prieto}, {Prosapio}, {Rivetta}, {Schneider}, {Snedden}, \&
  {Wang}}]{Gunn2006}
{Gunn}, J.~E., {Siegmund}, W.~A., {Mannery}, E.~J., {et~al.} 2006, \aj, 131,
  2332, \dodoi{10.1086/500975}

\bibitem[{{Gustafsson} {et~al.}(2008){Gustafsson}, {Edvardsson}, {Eriksson},
  {J{\o}rgensen}, {Nordlund}, \& {Plez}}]{Gustafsson2008}
{Gustafsson}, B., {Edvardsson}, B., {Eriksson}, K., {et~al.} 2008, \aap, 486,
  951, \dodoi{10.1051/0004-6361:200809724}

\bibitem[{{Heavens} {et~al.}(2004){Heavens}, {Panter}, {Jimenez}, \&
  {Dunlop}}]{heavens2004}
{Heavens}, A., {Panter}, B., {Jimenez}, R., \& {Dunlop}, J. 2004, \nat, 428,
  625, \dodoi{10.1038/nature02474}

\bibitem[{{Holtzman} {et~al.}(2018){Holtzman}, {Hasselquist}, {Shetrone},
  {Cunha}, {Allende Prieto}, {Anguiano}, {Bizyaev}, {Bovy}, {Casey},
  {Edvardsson}, {Johnson}, {J{\"o}nsson}, {Meszaros}, {Smith}, {Sobeck},
  {Zamora}, {Chojnowski}, {Fernandez-Trincado}, {Garcia-Hernandez}, {Majewski},
  {Pinsonneault}, {Souto}, {Stringfellow}, {Tayar}, {Troup}, \&
  {Zasowski}}]{Holtzman2018}
{Holtzman}, J.~A., {Hasselquist}, S., {Shetrone}, M., {et~al.} 2018, \aj, 156,
  125, \dodoi{10.3847/1538-3881/aad4f9}

\bibitem[{{Hosek Jr} {et~al.}(2020){Hosek Jr}, {Lu}, {Lam}, {Gautam},
  {Lockhart}, {Kim}, \& {Jia}}]{Hosek2020}
{Hosek Jr}, M.~W., {Lu}, J.~R., {Lam}, C.~Y., {et~al.} 2020, arXiv e-prints,
  arXiv:2006.06691.
\newblock \doarXiv{2006.06691}

\bibitem[{{Janz} {et~al.}(2016){Janz}, {Norris}, {Forbes}, {Huxor},
  {Romanowsky}, {Frank}, {Escudero}, {Faifer}, {Forte}, {Kannappan},
  {Maraston}, {Brodie}, {Strader}, \& {Thompson}}]{Janz2016}
{Janz}, J., {Norris}, M.~A., {Forbes}, D.~A., {et~al.} 2016, \mnras, 456, 617,
  \dodoi{10.1093/mnras/stv2636}

\bibitem[{{Johnston} {et~al.}(2020){Johnston}, {Puzia}, {D'Ago}, {Eigenthaler},
  {Galaz}, {H{\"a}u{\ss}ler}, {Mora}, {Ordenes-Brice{\~n}o}, {Rong},
  {Spengler}, {Vogt}, {C{\^o}t{\'e}}, {Grebel}, {Hilker}, {Mieske}, {Miller},
  {S{\'a}nchez-Janssen}, {Taylor}, \& {Zhang}}]{JohnstonEvelyn2020}
{Johnston}, E.~J., {Puzia}, T.~H., {D'Ago}, G., {et~al.} 2020, \mnras,
  \dodoi{10.1093/mnras/staa1261}

\bibitem[{{J{\"o}nsson} {et~al.}(2018){J{\"o}nsson}, {Allende Prieto},
  {Holtzman}, {Feuillet}, {Hawkins}, {Cunha}, {M{\'e}sz{\'a}ros},
  {Hasselquist}, {Fern{\'a}ndez-Trincado}, {Garc{\'\i}a-Hern{\'a}ndez},
  {Bizyaev}, {Carrera}, {Majewski}, {Pinsonneault}, {Shetrone}, {Smith},
  {Sobeck}, {Souto}, {Stringfellow}, {Teske}, \& {Zamora}}]{jonsson2018}
{J{\"o}nsson}, H., {Allende Prieto}, C., {Holtzman}, J.~A., {et~al.} 2018, \aj,
  156, 126, \dodoi{10.3847/1538-3881/aad4f5}

\bibitem[{{J{\"o}nsson} {et~al.}(2020){J{\"o}nsson}, {Holtzman}, {Prieto},
  {Cunha}, {Garc{\'\i}a-Hern{\'a}ndez}, {Hasselquist}, {Masseron}, {Osorio},
  {Shetrone}, {Smith}, {Stringfellow}, {Bizyaev}, {Edvardsson}, {Majewski},
  {M{\'e}sz{\'a}ros}, {Souto}, {Zamora}, {Beaton}, {Bovy}, {Donor},
  {Pinsonneault}, {Poovelil}, \& {Sobeck}}]{jonsson2020}
{J{\"o}nsson}, H., {Holtzman}, J.~A., {Prieto}, C.~A., {et~al.} 2020, \aj, 160,
  120, \dodoi{10.3847/1538-3881/aba592}

\bibitem[{{Koleva} {et~al.}(2009){Koleva}, {Prugniel}, {Bouchard}, \&
  {Wu}}]{Koleva2009}
{Koleva}, M., {Prugniel}, P., {Bouchard}, A., \& {Wu}, Y. 2009, \aap, 501,
  1269, \dodoi{10.1051/0004-6361/200811467}

\bibitem[{{Kroupa}(2001)}]{Kroupa2001}
{Kroupa}, P. 2001, \mnras, 322, 231, \dodoi{10.1046/j.1365-8711.2001.04022.x}

\bibitem[{{La Barbera} {et~al.}(2017){La Barbera}, {Vazdekis}, {Ferreras},
  {Pasquali}, {Allende Prieto}, {R{\"o}ck}, {Aguado}, \&
  {Peletier}}]{labarbera2017}
{La Barbera}, F., {Vazdekis}, A., {Ferreras}, I., {et~al.} 2017, \mnras, 464,
  3597, \dodoi{10.1093/mnras/stw2407}

\bibitem[{{La Barbera} {et~al.}(2016){La Barbera}, {Vazdekis}, {Ferreras},
  {Pasquali}, {Cappellari}, {Mart{\'\i}n-Navarro}, {Sch{\"o}nebeck}, \&
  {Falc{\'o}n-Barroso}}]{labarbera2016}
---. 2016, \mnras, 457, 1468, \dodoi{10.1093/mnras/stv2996}

\bibitem[{Leitherer {et~al.}(2014)Leitherer, Ekstr{\"o}m, Meynet, Schaerer,
  Agienko, \& Levesque}]{Leitherer2014}
Leitherer, C., Ekstr{\"o}m, S., Meynet, G., {et~al.} 2014, The Astrophysical
  Journal Supplement Series, 212, 14, \dodoi{10.1088/0067-0049/212/1/14}

\bibitem[{{Leung} \& {Bovy}(2019)}]{Leung2019}
{Leung}, H.~W., \& {Bovy}, J. 2019, \mnras, 483, 3255,
  \dodoi{10.1093/mnras/sty3217}

\bibitem[{{Lewis} {et~al.}(2015){Lewis}, {Dalcanton}, {Dolphin}, {Weisz}, \&
  {Williams}}]{Lewis2015}
{Lewis}, A.~R., {Dalcanton}, J.~J., {Dolphin}, A.~E., {Weisz}, D.~R., \&
  {Williams}, B.~F. 2015, in IAU Symposium, Vol. 309, Galaxies in 3D across the
  Universe, ed. B.~L. {Ziegler}, F.~{Combes}, H.~{Dannerbauer}, \&
  M.~{Verdugo}, 57--60

\bibitem[{{Lu} {et~al.}(2013){Lu}, {Do}, {Ghez}, {Morris}, {Yelda}, \&
  {Matthews}}]{Lu2013}
{Lu}, J.~R., {Do}, T., {Ghez}, A.~M., {et~al.} 2013, \apj, 764, 155,
  \dodoi{10.1088/0004-637X/764/2/155}

\bibitem[{{MacArthur} {et~al.}(2010){MacArthur}, {McDonald}, {Courteau}, \&
  {Jes{\'u}s Gonz{\'a}lez}}]{MacArthur2010}
{MacArthur}, L.~A., {McDonald}, M., {Courteau}, S., \& {Jes{\'u}s
  Gonz{\'a}lez}, J. 2010, \apj, 718, 768, \dodoi{10.1088/0004-637X/718/2/768}

\bibitem[{{Mackereth} {et~al.}(2019){Mackereth}, {Bovy}, {Leung}, {Schiavon},
  {Trick}, {Chaplin}, {Cunha}, {Feuillet}, {Majewski}, {Martig}, {Miglio},
  {Nidever}, {Pinsonneault}, {Aguirre}, {Sobeck}, {Tayar}, \&
  {Zasowski}}]{Mackereth2019}
{Mackereth}, J.~T., {Bovy}, J., {Leung}, H.~W., {et~al.} 2019, \mnras, 489,
  176, \dodoi{10.1093/mnras/stz1521}

\bibitem[{{Majewski} {et~al.}(2017){Majewski}, {Schiavon}, {Frinchaboy},
  {Allende Prieto}, {Barkhouser}, {Bizyaev}, {Blank}, {Brunner}, {Burton},
  {Carrera}, {Chojnowski}, {Cunha}, {Epstein}, {Fitzgerald}, {Garc{\'\i}a
  P{\'e}rez}, {Hearty}, {Henderson}, {Holtzman}, {Johnson}, {Lam}, {Lawler},
  {Maseman}, {M{\'e}sz{\'a}ros}, {Nelson}, {Nguyen}, {Nidever}, {Pinsonneault},
  {Shetrone}, {Smee}, {Smith}, {Stolberg}, {Skrutskie}, {Walker}, {Wilson},
  {Zasowski}, {Anders}, {Basu}, {Beland}, {Blanton}, {Bovy}, {Brownstein},
  {Carlberg}, {Chaplin}, {Chiappini}, {Eisenstein}, {Elsworth}, {Feuillet},
  {Fleming}, {Galbraith-Frew}, {Garc{\'\i}a}, {Garc{\'\i}a-Hern{\'a}ndez},
  {Gillespie}, {Girardi}, {Gunn}, {Hasselquist}, {Hayden}, {Hekker}, {Ivans},
  {Kinemuchi}, {Klaene}, {Mahadevan}, {Mathur}, {Mosser}, {Muna}, {Munn},
  {Nichol}, {O'Connell}, {Parejko}, {Robin}, {Rocha-Pinto}, {Schultheis},
  {Serenelli}, {Shane}, {Silva Aguirre}, {Sobeck}, {Thompson}, {Troup},
  {Weinberg}, \& {Zamora}}]{Majewski2017}
{Majewski}, S.~R., {Schiavon}, R.~P., {Frinchaboy}, P.~M., {et~al.} 2017, \aj,
  154, 94, \dodoi{10.3847/1538-3881/aa784d}

\bibitem[{{Maraston}(2005)}]{Maraston2005}
{Maraston}, C. 2005, \mnras, 362, 799, \dodoi{10.1111/j.1365-2966.2005.09270.x}

\bibitem[{{Maraston} \& {Str{\"o}mb{\"a}ck}(2011)}]{Maraston2011}
{Maraston}, C., \& {Str{\"o}mb{\"a}ck}, G. 2011, \mnras, 418, 2785,
  \dodoi{10.1111/j.1365-2966.2011.19738.x}

\bibitem[{{Maraston} {et~al.}(2009){Maraston}, {Str{\"o}mb{\"a}ck}, {Thomas},
  {Wake}, \& {Nichol}}]{Maraston2009}
{Maraston}, C., {Str{\"o}mb{\"a}ck}, G., {Thomas}, D., {Wake}, D.~A., \&
  {Nichol}, R.~C. 2009, \mnras, 394, L107,
  \dodoi{10.1111/j.1745-3933.2009.00621.x}

\bibitem[{{Marigo} {et~al.}(2017){Marigo}, {Girardi}, {Bressan}, {Rosenfield},
  {Aringer}, {Chen}, {Dussin}, {Nanni}, {Pastorelli}, {Rodrigues}, {Trabucchi},
  {Bladh}, {Dalcanton}, {Groenewegen}, {Montalb{\'a}n}, \& {Wood}}]{Marigo2017}
{Marigo}, P., {Girardi}, L., {Bressan}, A., {et~al.} 2017, \apj, 835, 77,
  \dodoi{10.3847/1538-4357/835/1/77}

\bibitem[{{M{\'a}rmol-Queralt{\'o}} {et~al.}(2008){M{\'a}rmol-Queralt{\'o}},
  {Cardiel}, {Cenarro}, {Vazdekis}, {Gorgas}, {Pedraz}, {Peletier}, \&
  {S{\'a}nchez-Bl{\'a}zquez}}]{marmol2008}
{M{\'a}rmol-Queralt{\'o}}, E., {Cardiel}, N., {Cenarro}, A.~J., {et~al.} 2008,
  \aap, 489, 885, \dodoi{10.1051/0004-6361:200810044}

\bibitem[{{Martins} {et~al.}(2019){Martins}, {Lima-Dias}, {Coelho}, \&
  {Lagan{\'a}}}]{Martins2019}
{Martins}, L.~P., {Lima-Dias}, C., {Coelho}, P. R.~T., \& {Lagan{\'a}}, T.~F.
  2019, \mnras, 484, 2388, \dodoi{10.1093/mnras/stz126}

\bibitem[{{McDermid} {et~al.}(2015){McDermid}, {Alatalo}, {Blitz}, {Bournaud},
  {Bureau}, {Cappellari}, {Crocker}, {Davies}, {Davis}, {de Zeeuw}, {Duc},
  {Emsellem}, {Khochfar}, {Krajnovi{\'c}}, {Kuntschner}, {Morganti}, {Naab},
  {Oosterloo}, {Sarzi}, {Scott}, {Serra}, {Weijmans}, \&
  {Young}}]{McDermid2015}
{McDermid}, R.~M., {Alatalo}, K., {Blitz}, L., {et~al.} 2015, \mnras, 448,
  3484, \dodoi{10.1093/mnras/stv105}

\bibitem[{{McGregor} {et~al.}(2003){McGregor}, {Hart}, {Conroy}, {Pfitzner},
  {Bloxham}, {Jones}, {Downing}, {Dawson}, {Young}, {Jarnyk}, \& {Van
  Harmelen}}]{NIFS}
{McGregor}, P.~J., {Hart}, J., {Conroy}, P.~G., {et~al.} 2003, Society of
  Photo-Optical Instrumentation Engineers (SPIE) Conference Series, Vol. 4841,
  {Gemini near-infrared integral field spectrograph (NIFS)}, ed. M.~{Iye} \&
  A.~F.~M. {Moorwood}, 1581--1591

\bibitem[{{Meneses-Goytia} {et~al.}(2015){Meneses-Goytia}, {Peletier},
  {Trager}, \& {Vazdekis}}]{Meneses-Goytia2015}
{Meneses-Goytia}, S., {Peletier}, R.~F., {Trager}, S.~C., \& {Vazdekis}, A.
  2015, \aap, 582, A97, \dodoi{10.1051/0004-6361/201423838}

\bibitem[{{Mould}(1978)}]{Mould1978}
{Mould}, J.~R. 1978, \apj, 220, 434, \dodoi{10.1086/155922}

\bibitem[{{Nidever} {et~al.}(2015){Nidever}, {Holtzman}, {Allende Prieto},
  {Beland}, {Bender}, {Bizyaev}, {Burton}, {Desphande}, {Fleming}, {Garc{\'\i}a
  P{\'e}rez}, {Hearty}, {Majewski}, {M{\'e}sz{\'a}ros}, {Muna}, {Nguyen},
  {Schiavon}, {Shetrone}, {Skrutskie}, {Sobeck}, \& {Wilson}}]{Nidever2015}
{Nidever}, D.~L., {Holtzman}, J.~A., {Allende Prieto}, C., {et~al.} 2015, \aj,
  150, 173, \dodoi{10.1088/0004-6256/150/6/173}

\bibitem[{{Ocvirk} {et~al.}(2006){Ocvirk}, {Pichon}, {Lan{\c{c}}on}, \&
  {Thi{\'e}baut}}]{ocvirk2006}
{Ocvirk}, P., {Pichon}, C., {Lan{\c{c}}on}, A., \& {Thi{\'e}baut}, E. 2006,
  \mnras, 365, 46, \dodoi{10.1111/j.1365-2966.2005.09182.x}

\bibitem[{{Onken} {et~al.}(2014){Onken}, {Valluri}, {Brown}, {McGregor},
  {Peterson}, {Bentz}, {Ferrarese}, {Pogge}, {Vestergaard}, {Storchi-Bergmann},
  \& {Riffel}}]{Onken2014}
{Onken}, C.~A., {Valluri}, M., {Brown}, J.~S., {et~al.} 2014, \apj, 791, 37,
  \dodoi{10.1088/0004-637X/791/1/37}

\bibitem[{{Pace} {et~al.}(2019){Pace}, {Tremonti}, {Chen}, {Schaefer},
  {Bershady}, {Westfall}, {Boquien}, {Rowlands}, {Andrews}, {Brownstein},
  {Drory}, \& {Wake}}]{Pace2019}
{Pace}, Z.~J., {Tremonti}, C., {Chen}, Y., {et~al.} 2019, \apj, 883, 82,
  \dodoi{10.3847/1538-4357/ab3723}

\bibitem[{{Parikh} {et~al.}(2019){Parikh}, {Thomas}, {Maraston}, {Westfall},
  {Lian}, {Fraser-McKelvie}, {Andrews}, {Drory}, \&
  {Meneses-Goytia}}]{Parikh2019}
{Parikh}, T., {Thomas}, D., {Maraston}, C., {et~al.} 2019, \mnras, 483, 3420,
  \dodoi{10.1093/mnras/sty3339}

\bibitem[{{Pastorelli} {et~al.}(2019){Pastorelli}, {Marigo}, {Girardi}, {Chen},
  {Rubele}, {Trabucchi}, {Aringer}, {Bladh}, {Bressan}, {Montalb{\'a}n},
  {Boyer}, {Dalcanton}, {Eriksson}, {Groenewegen}, {H{\"o}fner}, {Lebzelter},
  {Nanni}, {Rosenfield}, {Wood}, \& {Cioni}}]{Pastorelli2019}
{Pastorelli}, G., {Marigo}, P., {Girardi}, L., {et~al.} 2019, \mnras, 485,
  5666, \dodoi{10.1093/mnras/stz725}

\bibitem[{{Paxton} {et~al.}(2011){Paxton}, {Bildsten}, {Dotter}, {Herwig},
  {Lesaffre}, \& {Timmes}}]{Paxton2011}
{Paxton}, B., {Bildsten}, L., {Dotter}, A., {et~al.} 2011, \apjs, 192, 3,
  \dodoi{10.1088/0067-0049/192/1/3}

\bibitem[{{Paxton} {et~al.}(2013){Paxton}, {Cantiello}, {Arras}, {Bildsten},
  {Brown}, {Dotter}, {Mankovich}, {Montgomery}, {Stello}, {Timmes}, \&
  {Townsend}}]{Paxton2013}
{Paxton}, B., {Cantiello}, M., {Arras}, P., {et~al.} 2013, \apjs, 208, 4,
  \dodoi{10.1088/0067-0049/208/1/4}

\bibitem[{{Paxton} {et~al.}(2015){Paxton}, {Marchant}, {Schwab}, {Bauer},
  {Bildsten}, {Cantiello}, {Dessart}, {Farmer}, {Hu}, {Langer}, {Townsend},
  {Townsley}, \& {Timmes}}]{Paxton2015}
{Paxton}, B., {Marchant}, P., {Schwab}, J., {et~al.} 2015, \apjs, 220, 15,
  \dodoi{10.1088/0067-0049/220/1/15}

\bibitem[{{P{\'e}rez} {et~al.}(2013){P{\'e}rez}, {Cid Fernandes}, {Gonz{\'a}lez
  Delgado}, {Garc{\'\i}a-Benito}, {S{\'a}nchez}, {Husemann}, {Mast},
  {Rod{\'o}n}, {Kupko}, {Backsmann}, {de Amorim}, {van de Ven}, {Walcher},
  {Wisotzki}, {Cortijo-Ferrero}, \& {CALIFA Collaboration}}]{Perez2013}
{P{\'e}rez}, E., {Cid Fernandes}, R., {Gonz{\'a}lez Delgado}, R.~M., {et~al.}
  2013, \apjl, 764, L1, \dodoi{10.1088/2041-8205/764/1/L1}

\bibitem[{{Pickles}(1998)}]{Pickles1998}
{Pickles}, A.~J. 1998, \pasp, 110, 863, \dodoi{10.1086/316197}

\bibitem[{{Prugniel} {et~al.}(2007){Prugniel}, {Soubiran}, {Koleva}, \& {Le
  Borgne}}]{Prugniel2007}
{Prugniel}, P., {Soubiran}, C., {Koleva}, M., \& {Le Borgne}, D. 2007, arXiv
  e-prints, astro.
\newblock \doarXiv{astro-ph/0703658}

\bibitem[{{Rayner} {et~al.}(2009){Rayner}, {Cushing}, \& {Vacca}}]{Rayner2009}
{Rayner}, J.~T., {Cushing}, M.~C., \& {Vacca}, W.~D. 2009, \apjs, 185, 289,
  \dodoi{10.1088/0067-0049/185/2/289}

\bibitem[{{Riffel} {et~al.}(2019){Riffel}, {Rodr{\'\i}guez-Ardila},
  {Brotherton}, {Peletier}, {Vazdekis}, {Riffel}, {Martins}, {Bonatto}, {Zanon
  Dametto}, {Dahmer-Hahn}, {Runnoe}, {Pastoriza}, {Chies-Santos}, \&
  {Trevisan}}]{Riffel2019}
{Riffel}, R., {Rodr{\'\i}guez-Ardila}, A., {Brotherton}, M.~S., {et~al.} 2019,
  \mnras, 486, 3228, \dodoi{10.1093/mnras/stz1077}

\bibitem[{{R{\"o}ck} {et~al.}(2017){R{\"o}ck}, {Vazdekis}, {La Barbera},
  {Peletier}, {Knapen}, {Allende-Prieto}, \& {Aguado}}]{rock2017}
{R{\"o}ck}, B., {Vazdekis}, A., {La Barbera}, F., {et~al.} 2017, \mnras, 472,
  361, \dodoi{10.1093/mnras/stx1940}

\bibitem[{{R{\"o}ck} {et~al.}(2016){R{\"o}ck}, {Vazdekis}, {Ricciardelli},
  {Peletier}, {Knapen}, \& {Falc{\'o}n-Barroso}}]{Rock2016}
{R{\"o}ck}, B., {Vazdekis}, A., {Ricciardelli}, E., {et~al.} 2016, \aap, 589,
  A73, \dodoi{10.1051/0004-6361/201527570}

\bibitem[{{Ruiz-Lara} {et~al.}(2020){Ruiz-Lara}, {Gallart}, {Bernard}, \&
  {Cassisi}}]{Ruiz-lara2020}
{Ruiz-Lara}, T., {Gallart}, C., {Bernard}, E.~J., \& {Cassisi}, S. 2020, arXiv
  e-prints, arXiv:2003.12577.
\newblock \doarXiv{2003.12577}

\bibitem[{{Sakari} {et~al.}(2016){Sakari}, {Shetrone}, {Schiavon}, {Bizyaev},
  {Prieto}, {Beers}, {Caldwell}, {Garcia-Hernandez}, {Lucatello}, {Majewski},
  {O'Connell}, {Pan}, \& {Strader}}]{Sakari2016}
{Sakari}, C.~M., {Shetrone}, M.~D., {Schiavon}, R.~P., {et~al.} 2016, VizieR
  Online Data Catalog, J/ApJ/829/116

\bibitem[{{Salaris} \& {Cassisi}(2005)}]{Salaris2005}
{Salaris}, M., \& {Cassisi}, S. 2005, {Evolution of Stars and Stellar
  Populations}

\bibitem[{{S{\'a}nchez} {et~al.}(2016{\natexlab{a}}){S{\'a}nchez}, {P{\'e}rez},
  {S{\'a}nchez-Bl{\'a}zquez}, {Gonz{\'a}lez}, {Ros{\'a}les-Ortega},
  {Cano-D{\'\i}az}, {L{\'o}pez-Cob{\'a}}, {Marino}, {Gil de Paz}, {Moll{\'a}},
  {L{\'o}pez-S{\'a}nchez}, {Ascasibar}, \&
  {Barrera-Ballesteros}}]{sanchez2016a}
{S{\'a}nchez}, S.~F., {P{\'e}rez}, E., {S{\'a}nchez-Bl{\'a}zquez}, P., {et~al.}
  2016{\natexlab{a}}, \rmxaa, 52, 21.
\newblock \doarXiv{1509.08552}

\bibitem[{{S{\'a}nchez} {et~al.}(2016{\natexlab{b}}){S{\'a}nchez}, {P{\'e}rez},
  {S{\'a}nchez-Bl{\'a}zquez}, {Garc{\'\i}a-Benito}, {Ibarra-Mede},
  {Gonz{\'a}lez}, {Rosales-Ortega}, {S{\'a}nchez-Menguiano}, {Ascasibar},
  {Bitsakis}, {Law}, {Cano-D{\'\i}az}, {L{\'o}pez-Cob{\'a}}, {Marino}, {Gil de
  Paz}, {L{\'o}pez-S{\'a}nchez}, {Barrera-Ballesteros}, {Galbany}, {Mast},
  {Abril-Melgarejo}, \& {Roman-Lopes}}]{sanchez2016b}
---. 2016{\natexlab{b}}, \rmxaa, 52, 171.
\newblock \doarXiv{1602.01830}

\bibitem[{{Schiavon}(2007)}]{Schiavon2007}
{Schiavon}, R.~P. 2007, \apjs, 171, 146, \dodoi{10.1086/511753}

\bibitem[{{Schiavon} {et~al.}(2012){Schiavon}, {Caldwell}, {Morrison},
  {Harding}, {Courteau}, {MacArthur}, \& {Graves}}]{Schiavon2012}
{Schiavon}, R.~P., {Caldwell}, N., {Morrison}, H., {et~al.} 2012, \aj, 143, 14,
  \dodoi{10.1088/0004-6256/143/1/14}

\bibitem[{{Serenelli} {et~al.}(2017){Serenelli}, {Weiss}, {Cassisi}, {Salaris},
  \& {Pietrinferni}}]{Serenelli2017}
{Serenelli}, A., {Weiss}, A., {Cassisi}, S., {Salaris}, M., \& {Pietrinferni},
  A. 2017, \aap, 606, A33, \dodoi{10.1051/0004-6361/201731004}

\bibitem[{{Silva} {et~al.}(2008){Silva}, {Kuntschner}, \&
  {Lyubenova}}]{silva2008}
{Silva}, D.~R., {Kuntschner}, H., \& {Lyubenova}, M. 2008, \apj, 674, 194,
  \dodoi{10.1086/524869}

\bibitem[{{Spinrad} \& {Taylor}(1971)}]{Spinrad1971}
{Spinrad}, H., \& {Taylor}, B.~J. 1971, \apjs, 22, 445, \dodoi{10.1086/190232}

\bibitem[{{Steidel} {et~al.}(2017){Steidel}, {Rudie}, {Strom}, {Pettini},
  {Reddy}, {Shapley}, {Trainor}, {Erb}, {Turner}, {Konidaris}, {Kulas}, {Mace},
  {Matthews}, \& {McLean}}]{MOSFIRE}
{Steidel}, C.~C., {Rudie}, G.~C., {Strom}, A.~L., {et~al.} 2017, VizieR Online
  Data Catalog, J/ApJ/795/165

\bibitem[{{Strader} {et~al.}(2011){Strader}, {Caldwell}, \&
  {Seth}}]{Strader2011}
{Strader}, J., {Caldwell}, N., \& {Seth}, A.~C. 2011, \aj, 142, 8,
  \dodoi{10.1088/0004-6256/142/1/8}

\bibitem[{{Tang} {et~al.}(2014){Tang}, {Bressan}, {Rosenfield}, {Slemer},
  {Marigo}, {Girardi}, \& {Bianchi}}]{Tang2014}
{Tang}, J., {Bressan}, A., {Rosenfield}, P., {et~al.} 2014, \mnras, 445, 4287,
  \dodoi{10.1093/mnras/stu2029}

\bibitem[{{Thomas} {et~al.}(2003){Thomas}, {Maraston}, \&
  {Bender}}]{Thomas2003}
{Thomas}, D., {Maraston}, C., \& {Bender}, R. 2003, \mnras, 339, 897,
  \dodoi{10.1046/j.1365-8711.2003.06248.x}

\bibitem[{{Tinsley}(1968)}]{Tinsley1968}
{Tinsley}, B.~M. 1968, \apj, 151, 547, \dodoi{10.1086/149455}

\bibitem[{{Tojeiro}(2007)}]{tojeiro2007}
{Tojeiro}, R. 2007, in Astronomical Society of the Pacific Conference Series,
  Vol. 374, From Stars to Galaxies: Building the Pieces to Build Up the
  Universe, ed. A.~{Vallenari}, R.~{Tantalo}, L.~{Portinari}, \& A.~{Moretti},
  507

\bibitem[{{Trager} {et~al.}(2000){Trager}, {Faber}, {Worthey}, \&
  {Gonz{\'a}lez}}]{Trager2000}
{Trager}, S.~C., {Faber}, S.~M., {Worthey}, G., \& {Gonz{\'a}lez}, J.~J. 2000,
  \aj, 119, 1645, \dodoi{10.1086/301299}

\bibitem[{{Vazdekis} {et~al.}(1996){Vazdekis}, {Casuso}, {Peletier}, \&
  {Beckman}}]{Vazdekis1996}
{Vazdekis}, A., {Casuso}, E., {Peletier}, R.~F., \& {Beckman}, J.~E. 1996,
  \apjs, 106, 307, \dodoi{10.1086/192340}

\bibitem[{{Vazdekis} {et~al.}(2016){Vazdekis}, {Koleva}, {Ricciardelli},
  {R{\"o}ck}, \& {Falc{\'o}n-Barroso}}]{vazdekis2016}
{Vazdekis}, A., {Koleva}, M., {Ricciardelli}, E., {R{\"o}ck}, B., \&
  {Falc{\'o}n-Barroso}, J. 2016, \mnras, 463, 3409,
  \dodoi{10.1093/mnras/stw2231}

\bibitem[{{Vazdekis} {et~al.}(2012){Vazdekis}, {Ricciardelli}, {Cenarro},
  {Rivero-Gonz{\'a}lez}, {D{\'\i}az-Garc{\'\i}a}, \&
  {Falc{\'o}n-Barroso}}]{Vazdekis2012}
{Vazdekis}, A., {Ricciardelli}, E., {Cenarro}, A.~J., {et~al.} 2012, \mnras,
  424, 157, \dodoi{10.1111/j.1365-2966.2012.21179.x}

\bibitem[{{Vazdekis} {et~al.}(2010){Vazdekis}, {S{\'a}nchez-Bl{\'a}zquez},
  {Falc{\'o}n-Barroso}, {Cenarro}, {Beasley}, {Cardiel}, {Gorgas}, \&
  {Peletier}}]{Vazdekis2010}
{Vazdekis}, A., {S{\'a}nchez-Bl{\'a}zquez}, P., {Falc{\'o}n-Barroso}, J.,
  {et~al.} 2010, \mnras, 404, 1639, \dodoi{10.1111/j.1365-2966.2010.16407.x}

\bibitem[{{Weisz} {et~al.}(2011){Weisz}, {Dolphin}, {Dalcanton}, {Skillman},
  {Holtzman}, {Williams}, {Gilbert}, {Seth}, {Cole}, {Gogarten}, {Rosema},
  {Karachentsev}, {McQuinn}, \& {Zaritsky}}]{Weisz2011}
{Weisz}, D.~R., {Dolphin}, A.~E., {Dalcanton}, J.~J., {et~al.} 2011, \apj, 743,
  8, \dodoi{10.1088/0004-637X/743/1/8}

\bibitem[{{Wilkinson} {et~al.}(2017){Wilkinson}, {Maraston}, {Goddard},
  {Thomas}, \& {Parikh}}]{wilkinson2017}
{Wilkinson}, D.~M., {Maraston}, C., {Goddard}, D., {Thomas}, D., \& {Parikh},
  T. 2017, \mnras, 472, 4297, \dodoi{10.1093/mnras/stx2215}

\bibitem[{{Williams} {et~al.}(2009){Williams}, {Dalcanton}, {Seth}, {Weisz},
  {Dolphin}, {Skillman}, {Harris}, {Holtzman}, {Girardi}, {de Jong}, {Olsen},
  {Cole}, {Gallart}, {Gogarten}, {Hidalgo}, {Mateo}, {Rosema}, {Stetson}, \&
  {Quinn}}]{Williams2009}
{Williams}, B.~F., {Dalcanton}, J.~J., {Seth}, A.~C., {et~al.} 2009, \aj, 137,
  419, \dodoi{10.1088/0004-6256/137/1/419}

\bibitem[{{Wilson} {et~al.}(2019){Wilson}, {Hearty}, {Skrutskie}, {Majewski},
  {Holtzman}, {Eisenstein}, {Gunn}, {Blank}, {Henderson}, {Smee}, {Nelson},
  {Nidever}, {Arns}, {Barkhouser}, {Barr}, {Beland}, {Bershady}, {Blanton},
  {Brunner}, {Burton}, {Carey}, {Carr}, {Colque}, {Crane}, {Damke}, {Davidson},
  {Dean}, {Di Mille}, {Don}, {Ebelke}, {Evans}, {Fitzgerald}, {Gillespie},
  {Hall}, {Harding}, {Harding}, {Hammond}, {Hancock}, {Harrison}, {Hope},
  {Horne}, {Karakla}, {Lam}, {Leger}, {MacDonald}, {Maseman}, {Matsunari},
  {Melton}, {Mitcheltree}, {O'Brien}, {O'Connell}, {Patten}, {Richardson},
  {Rieke}, {Rieke}, {Roman-Lopes}, {Schiavon}, {Sobeck}, {Stolberg}, {Stoll},
  {Tembe}, {Trujillo}, {Uomoto}, {Vernieri}, {Walker}, {Weinberg}, {Young},
  {Anthony-Brumfield}, {Bizyaev}, {Breslauer}, {De Lee}, {Downey}, {Halverson},
  {Huehnerhoff}, {Klaene}, {Leon}, {Long}, {Mahadevan}, {Malanushenko},
  {Nguyen}, {Owen}, {S{\'a}nchez-Gallego}, {Sayres}, {Shane}, {Shectman},
  {Shetrone}, {Skinner}, {Stauffer}, \& {Zhao}}]{Wilson2019}
{Wilson}, J.~C., {Hearty}, F.~R., {Skrutskie}, M.~F., {et~al.} 2019, \pasp,
  131, 055001, \dodoi{10.1088/1538-3873/ab0075}

\bibitem[{{Wisnioski} {et~al.}(2015){Wisnioski}, {F{\"o}rster Schreiber},
  {Wuyts}, {Wuyts}, {Bandara}, {Wilman}, {Genzel}, {Bender}, {Davies},
  {Fossati}, {Lang}, {Mendel}, {Beifiori}, {Brammer}, {Chan}, {Fabricius},
  {Fudamoto}, {Kulkarni}, {Kurk}, {Lutz}, {Nelson}, {Momcheva}, {Rosario},
  {Saglia}, {Seitz}, {Tacconi}, \& {van Dokkum}}]{KMOS}
{Wisnioski}, E., {F{\"o}rster Schreiber}, N.~M., {Wuyts}, S., {et~al.} 2015,
  \apj, 799, 209, \dodoi{10.1088/0004-637X/799/2/209}

\bibitem[{{Worthey} {et~al.}(1994){Worthey}, {Faber}, {Gonzalez}, \&
  {Burstein}}]{Worthey1994}
{Worthey}, G., {Faber}, S.~M., {Gonzalez}, J.~J., \& {Burstein}, D. 1994,
  \apjs, 94, 687, \dodoi{10.1086/192087}

\bibitem[{{Ygouf} {et~al.}(2017){Ygouf}, {Beichman}, {Hodapp}, \&
  {Roellig}}]{NIRSPEC}
{Ygouf}, M., {Beichman}, C., {Hodapp}, K., \& {Roellig}, T. 2017, in SF2A-2017:
  Proceedings of the Annual meeting of the French Society of Astronomy and
  Astrophysics, ed. C.~{Reyl{\'e}}, P.~{Di Matteo}, F.~{Herpin}, E.~{Lagadec},
  A.~{Lan{\c{c}}on}, Z.~{Meliani}, \& F.~{Royer}, Di

\bibitem[{{Zasowski} {et~al.}(2013){Zasowski}, {Johnson}, {Frinchaboy},
  {Majewski}, {Nidever}, {Rocha Pinto}, {Girardi}, {Andrews}, {Chojnowski},
  {Cudworth}, {Jackson}, {Munn}, {Skrutskie}, {Beaton}, {Blake}, {Covey},
  {Deshpande}, {Epstein}, {Fabbian}, {Fleming}, {Garcia Hernandez}, {Herrero},
  {Mahadevan}, {M{\'e}sz{\'a}ros}, {Schultheis}, {Sellgren}, {Terrien}, {van
  Saders}, {Allende Prieto}, {Bizyaev}, {Burton}, {Cunha}, {da Costa},
  {Hasselquist}, {Hearty}, {Holtzman}, {Garc{\'\i}a P{\'e}rez}, {Maia},
  {O'Connell}, {O'Donnell}, {Pinsonneault}, {Santiago}, {Schiavon}, {Shetrone},
  {Smith}, \& {Wilson}}]{Zasowski2013}
{Zasowski}, G., {Johnson}, J.~A., {Frinchaboy}, P.~M., {et~al.} 2013, \aj, 146,
  81, \dodoi{10.1088/0004-6256/146/4/81}

\bibitem[{{Zasowski} {et~al.}(2017){Zasowski}, {Cohen}, {Chojnowski},
  {Santana}, {Oelkers}, {Andrews}, {Beaton}, {Bender}, {Bird}, {Bovy},
  {Carlberg}, {Covey}, {Cunha}, {Dell'Agli}, {Fleming}, {Frinchaboy},
  {Garc{\'\i}a-Hern{\'a}ndez}, {Harding}, {Holtzman}, {Johnson}, {Kollmeier},
  {Majewski}, {M{\'e}sz{\'a}ros}, {Munn}, {Mu{\~n}oz}, {Ness}, {Nidever},
  {Poleski}, {Rom{\'a}n-Z{\'u}{\~n}iga}, {Shetrone}, {Simon}, {Smith},
  {Sobeck}, {Stringfellow}, {Szigeti{\'a}ros}, {Tayar}, \&
  {Troup}}]{Zasowski2017}
{Zasowski}, G., {Cohen}, R.~E., {Chojnowski}, S.~D., {et~al.} 2017, \aj, 154,
  198, \dodoi{10.3847/1538-3881/aa8df9}

\end{thebibliography}

\clearpage

\appendix
\section{Sample code to access A-LIST models}
\lstdefinestyle{mystyle}{
    backgroundcolor=\color{backcolour},   
    commentstyle=\color{codegreen},
    keywordstyle=\color{magenta},
    numberstyle=\tiny\color{codegray},
    stringstyle=\color{codepurple},
    basicstyle=\ttfamily\footnotesize,
    breakatwhitespace=true,         
    breaklines=true,                 
    captionpos=b,                    
    keepspaces=true,                 
    numbers=left,                    
    numbersep=5pt,                  
    showspaces=false,                
    showstringspaces=false,
    showtabs=false,                  
    tabsize=2
}
\lstset{style=mystyle}
\begin{lstlisting}[language=Python]
import numpy as np #(version: 1.18.1)
import matplotlib.pyplot as plt #(version: 2.1.0)
from astropy.table import Table, Column #(version: 4.0.1.post1)
from astropy.io import fits #(version: 4.0.1.post1)

#Reading the .fits file to access the age, [M/H] and [$\alpha$/M] as well as the spectral models.
#Here we are using astropy package in python.
sspgrid = Table.read('A-LIST_padova.fits') #Contains the table with age, [M/H] and [$\alpha$/M]
ssp_spec = fits.open('A-LIST_padova.fits')[2].data # contains the spectral models
ssp_spec_uncert = fits.open('ALIST_Padova_variance.fits')[2].data #contains the variance spectra for each model

##Selecting the models based on the quality cuts:
ixs_quality = np.where((sspgrid['lumfrac'] > 0.32) & (sspgrid['deltatemp'] > -200) & (sspgrid['deltatemp'] < 350))[0]

##Define the age, [M/H] and [$\alpha$/M] needed to read-in:
age, m_h, a_m = 10, 0.0, 0.0
##Pulling out a spectrum using the age, [M/H] and [$\alpha$/M] values from the table:
model_id = np.where((sspgrid[ixs_quality]['AGE'] == age) & (sspgrid[ixs_quality]['M_H'] == m_h) & \
                    (sspgrid[ixs_quality]['A_M'] == a_m))[0]
model_spec = ssp_spec[ixs_quality][model_id][0]

##Defining the wavelength range using the information provided in the header.
wavstart = float(fits.getheader('A-LIST_Padova.fits')['CRVAL1'].split()[0]) ##Starting wavelength in log scale
wavdelt = float(fits.getheader('A-LIST_Padova.fits')['CDELT1'].split()[0]) ##Delta wavelength in log space
numpix = int(fits.getheader('A-LIST_Padova.fits')['NWAVE'].split()[0]) ##Total number of pixels available in the spectra

wavend = wavstart + (wavdelt*numpix) ##The wavelength value for the last pixel
##Using the information avilable above, we create an array of wavelength in the linear space using numpy's linspace:
wavelength = 10**(np.linspace(wavstart, wavend, num = numpix))


##An example plot to show what the spectrum looks like:
plt.plot(wavelength, model_spec)
##These lines below are just the aesthetics of the plot.
plt.xlabel(r'$\lambda$($\AA$)')
plt.ylabel('flux')
plt.title(r'age = %d, [M/H] = %d, [$\alpha$/M] = %d'%(sspgrid['AGE'][model_id], sspgrid['M_H'][model_id],sspgrid['A_M'][model_id]))
plt.show()

\end{lstlisting}
\begin{figure*}[!b]
    \centering
    \includegraphics[clip, width=0.4\textwidth]{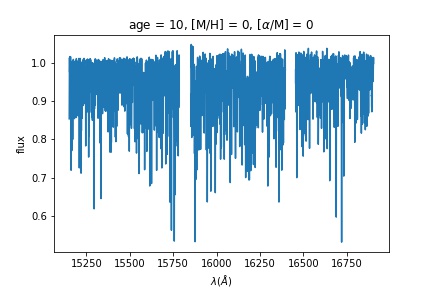}
\end{figure*}

\end{document}